\pdfoutput=1

\newif\ifdraft\draftfalse 
\newif\ifanon\anonfalse    
\newif\iffull\fullfalse    
\newif\iflongrefs\longrefsfalse 
\newif\ifbackref\backreffalse 
\newif\ifsooner\soonerfalse
\newif\iflater\laterfalse
\newif\ifcamera\cameratrue    
\newif\ifcheckpagebudget\checkpagebudgetfalse
\newif\ifconference\conferencetrue

\newcommand{\wand}{-\!\ast}

\makeatletter \@input{texdirectives.tex} \makeatother

\ifcamera
\documentclass[acmsmall,screen]{acmart}
\else
\ifanon
\documentclass[acmsmall,screen,review,anonymous]{acmart}
\else
\documentclass[acmsmall,review=false,screen=true]{acmart}
\fi
\fi

\ifanon\else
\setcopyright{rightsretained}
\acmPrice{}
\acmDOI{10.1145/3409003}
\acmYear{2020}
\copyrightyear{2020}
\acmSubmissionID{icfp20main-p102-p}
\acmJournal{PACMPL}
\acmVolume{4}
\acmNumber{ICFP}
\acmArticle{121}
\acmMonth{8}

\fi

\makeatletter
\def\@copyrightpermission{\ifcamera\\\\\\\fi This work is licensed under a \href{https://creativecommons.org/licenses/by/4.0/}{Creative Commons Attribution 4.0 International License}}
\makeatother

\ifcamera\else
\makeatletter
\def\@authorsaddresses{}
\def\@mkbibcitation{}
\makeatother
\fi

\usepackage{microtype}
\usepackage{afterpage}
\usepackage{color}
\usepackage{soul}
\usepackage{amsmath,amsfonts}
\usepackage{extarrows}
\usepackage{version}
\usepackage{xspace}
\usepackage{graphicx}
\usepackage{listings}
\usepackage{wrapfig}
\usepackage{ifpdf}
\usepackage{semantic}
\usepackage{mathpartir} 

\usepackage{natbib}
\newcommand\citepos[1]{\citeauthor{#1}'s\ (\citeyear{#1})}
\bibpunct();A{},
\let\cite=\citep
\usepackage{amsthm}
\usepackage{stmaryrd}
\usepackage{subfigure}
\usepackage{multirow}
\usepackage{multicol}
\usepackage{proof}
\usepackage{framed}
\usepackage{caption}
\usepackage{mathpartir}
\usepackage{enumitem}
\usepackage{tikz}
\usetikzlibrary{positioning}
\usetikzlibrary{arrows}
\usepackage[T1]{fontenc}
\usepackage[utf8]{inputenc}

\ifcamera
\usepackage{flushend} 
\fi
\definecolor{darkblue}{rgb}{0.0,0.0,0.3}

\usepackage{hyperref}
\hypersetup{breaklinks=true}

\newcommand\maybecolor[1]{\color{#1}}

\let\ls\lstinline

\def\Snospace~{\S{}}

\newcommand\fstar{F$^\star$\xspace}

\newcommand\steelcore{SteelCore\xspace}
\newcommand\steel{Steel\xspace}
\definecolor{addition}{rgb}{0,0.1,0.5}
\definecolor{dkblue}{rgb}{0,0.1,0.5}
\definecolor{dkgreen}{rgb}{0,0.4,0}
\definecolor{dkred}{rgb}{0.6,0,0}
\definecolor{dkpurple}{rgb}{0.7,0,1.0}
\definecolor{purple}{rgb}{0.9,0,1.0}
\definecolor{olive}{rgb}{0.4, 0.4, 0.0}
\definecolor{teal}{rgb}{0.0,0.4,0.4}
\definecolor{azure}{rgb}{0.0, 0.5, 1.0}
\definecolor{gray}{rgb}{0.5, 0.5, 0.5}
\definecolor{dkgrey}{rgb}{0.2, 0.2, 0.2}
\definecolor{lilac}{rgb}{0.70, 0.04, 0.08}

\newcommand{\comm}[3]{\ifcheckpagebudget\else\ifdraft{{\color{#1}[#2: #3]}}\fi\fi}

\newcommand{\nik}[1]{\comm{dkpurple}{Nik}{#1}}

\newcommand{\ch}[1]{\comm{lilac}{Catalin}{#1}}

\begin{document}
\title{SteelCore: An Extensible Concurrent Separation Logic
                  for Effectful Dependently Typed Programs}

\ifanon
\else
\author{Nikhil Swamy}
\affiliation{\institution{Microsoft Research}\country{USA}}
\author{Aseem Rastogi}
\affiliation{\institution{Microsoft Research}\country{India}}
\author{Aymeric Fromherz}
\affiliation{\institution{Carnegie Mellon University}\country{USA}}
\author{Denis Merigoux}
\affiliation{\institution{Inria Paris}\country{France}}  
\author{Danel Ahman}
\affiliation{\institution{University of Ljubljana}\country{Slovenia}}
\author{Guido Mart\'inez}
\affiliation{\institution{CIFASIS-CONICET}\country{Argentina}}
\makeatletter
\renewcommand{\@shortauthors}{N.~Swamy, A.~Rastogi, A.~Fromherz, D.~Merigoux, D.~Ahman, and G.~Mart\'inez}
\makeatother
\fi


\begin{abstract}
Much recent research has been devoted to modeling effects within type
theory. Building on this work, we observe that effectful type theories
can provide a foundation on which to build semantics for more complex
programming constructs and program logics, extending the reasoning
principles that apply within the host effectful type theory itself.

Concretely, our main contribution is a semantics for concurrent
separation logic (CSL) within the \fstar proof assistant in a manner
that enables dependently typed, effectful
\fstar programs to make use of concurrency and to be
specified and verified using a full-featured, extensible CSL.
In contrast to prior approaches, we directly derive the
partial-correctness Hoare rules for CSL from the denotation of
computations in the \emph{effectful} semantics of
non-deterministically interleaved atomic actions.

Demonstrating the flexibility of our semantics, we build generic,
verified libraries that support various concurrency constructs,
ranging from dynamically allocated, storable spin locks, to
protocol-indexed channels. We conclude that our effectful semantics
provides a simple yet expressive basis on which to layer
domain-specific languages and logics for verified, concurrent
programming.
\end{abstract}



\maketitle

\section{Introduction}
\label{sec:intro}

Proof assistants based on type theory can be a programmers'
delight, allowing one to build modular abstractions coupled with
strong specifications that ensure program correctness.
Their expressive power also allows one to develop new program logics
within the same framework as the programs
themselves.
A notable case in point is the Iris framework~\cite{JungKJBBD18}
embedded in Coq~\citep{coq}, which provides an impredicative,
higher-order, concurrent separation logic
(CSL)~\citep{reynolds02sep,ohearn04csl} within which to specify and
prove programs.

Iris has been used to model various languages and constructs, and to
verify many interesting programs~\cite{krogh2019aneris, Chajed:2019:VCC:3341301.3359632, hinrichsen19actris}.
However, Iris is not in itself a programming language: it must instead be
instantiated with a \emph{deeply embedded} representation
and semantics of one provided by the user.
For instance, several Iris-based papers work with a mini
ML-like language deeply embedded in Coq~\cite{krebbers17ipm}.

Taking a different approach, FCSL~\cite{nmb08htt, nanevski14fcsl,
nanevski19fcsl} embeds a predicative CSL in Coq enabling proofs of Coq
programs (rather than embedded-language programs) within a semantics
that accounts for effects like state and concurrency. This allows
programmers to use the full power of type theory not just for proving,
but also for programming, e.g., building dependently typed programs
and metaprograms over inductive datatypes, with typeclasses, a module
system, and other features of a full-fledged language. However,
Nanevski et al.'s program logics are inherently predicative, which
makes it difficult to express constructs like dynamically allocated
invariants and locks, which are natural in impredicative logics like
Iris.

In this paper, we develop a new framework called \steelcore that aims
to provide the benefits of Nanevski et al.'s shallow embeddings, while
also supporting dynamically allocated invariants and locks in the
flavor of Iris.
Specifically, we develop \steelcore in the \emph{effectful} type
theory provided by the \fstar proof assistant~\cite{mumon}.
One of our main insights is that an effectful type theory is not only
useful for programming; it can also be leveraged to build new program
logics for effectful program features like concurrency.
Building on prior work~\citep{preorders} that models the effect of
monotonic state in \fstar, we develop a semantics for concurrent
\fstar programs while simultaneously deriving a CSL to reason about \fstar
programs using the effect of concurrency.
The use of monotonic state enables us to account for invariants and
atomic actions entirely within \steelcore.
The net result is that we can program higher order,\ch{not spelled higher-order?}
dependently typed,
generally recursive, shared-memory and message-passing
concurrent \fstar programs and prove their partial correctness
using \steelcore.

\subsection{\steelcore: A Concurrent Separation Logic Embedded in \fstar}

\begin{figure}
\includegraphics[width=0.8\textwidth]{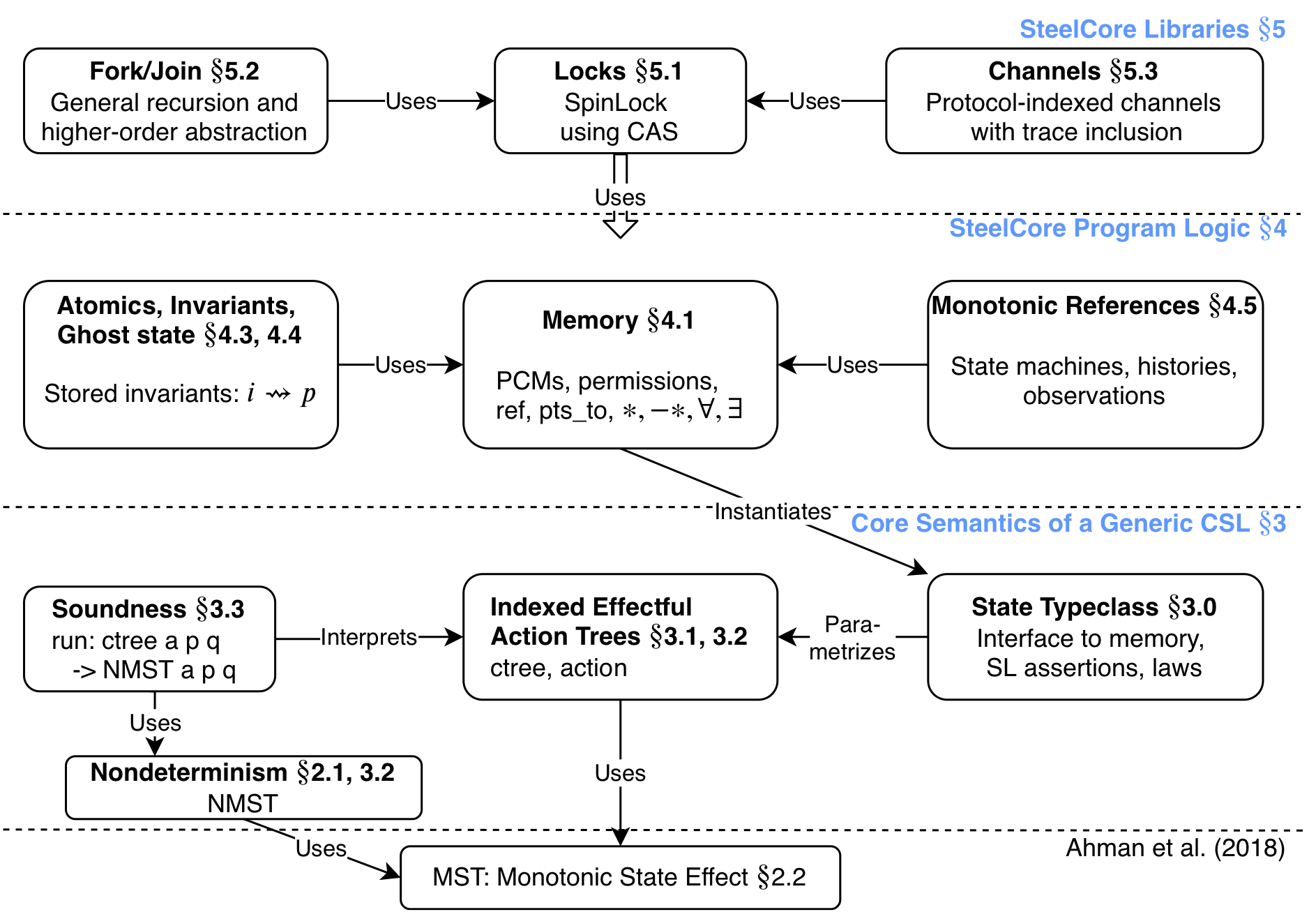}
\caption{An overview of \steelcore}
\label{fig:steelcore}
\end{figure}

\steelcore is the core semantics of \steel, a DSL under development in \fstar for
programming and proving concurrent programs. In this paper, we focus
primarily on the semantics, leaving a detailed treatment of
other aspects of the \steel framework to a separate paper.
The structure of \steelcore is shown in
Figure~\ref{fig:steelcore}. Building on the monotonic state effect, we
prove sound a generic program logic for concurrency, parametric in a
memory model and a separation logic (\S\ref{sec:semantics}). We
instantiate this semantics with a separation logic based on partial
commutative monoids, stored invariants, and state machines
(\S\ref{sec:steelcore}). Finally, using this logic, we program
verified, dependently typed, higher-order libraries for various kinds
of concurrency constructs, culminating in a library for message-passing
on typed channels (\S\ref{sec:examples}). We describe several novel
elements of our contributions, next.

For starters, we need to extend \fstar with concurrency. To do
this, we follow the well-known approach of encoding computational
effects as definitional interpreters over free
monads~\citep{hancock00interactive, swierstra2008alacarte,
kiselyov15freer, xia20interaction-trees}. That is, we can represent
computations as a datatype of (infinitely branching) trees of atomic
actions. When providing a computational interpretation for action
trees, one can pick an execution strategy (e.g., an interleaving
semantics) and build an interpreter to run programs. The first main
novelty of our work is that we provide an intrinsically typed
definitional interpreter~\citep{poulsen17definitional} that both
provides a semantics for concurrency while also deriving a CSL in
which to reason about concurrent programs. Enabling this development
is a new notion of indexed action trees, which we describe next.

\paragraph{Indexed action trees for structured parallelism}
We represent concurrent computations as an instance of the
datatype \ls`ctree st a pre post`, shown below. The \ls`ctree` type is
a tree of atomic computational actions, composed sequentially or in
parallel.
\begin{lstlisting}
type ctree (st:state) : a:Type -> pre:st.slprop -> post:(a -> st.slprop) -> Type =
 | Ret : x:a -> ctree st a (post x) post
 | Act : action pre post -> ctree st a pre post
 | Par : ctree st a p q -> ctree st a' p' q' -> ctree st (a & a') (p `st.star` p') (fun (x, x') -> q x `st.star` q' x')
 | Bind : ctree st a p q -> ((x:a) -> Dv (ctree st b (q x) r)) -> ctree st b p r
\end{lstlisting}
The type \ls`ctree st a pre post` is parameterized by an
instance \ls`st` of the \ls`state` typeclass, which provides a generic
interface to memories, including \ls`st.slprop`, the type of
separation logic assertions, and \ls`st.star`, the separating
conjunction. The index \ls`a` is the result type of the computation,
while \ls`pre` and \ls`post` are separation logic assertions.
The \ls`Act` nodes hold stateful atomic actions; \ls`Par` nodes
combine trees in parallel; while \ls`Bind` nodes sequentially compose
a computation with a potentially divergent continuation, as signified
by the \ls`Dv` effect label. Divergent computations are primitively
expressible
within \fstar, and are soundly isolated from its logical core of
total functions by the effect system.

\paragraph{Interpreting action trees in the effects of nondeterminism and monotonic state}
We interpret a term \ls`(e : ctree st a pre post)` as both a
computation \ls`e` as well as a proof of its own partial correctness
Hoare triple
\ls`{pre} e : a {post}`.
To prove this sound, we define an interpreter that
non-deterministically interleaves atomic actions run in parallel. The
interpreter is itself an effectful \fstar function with the following
(simplified) type, capturing our main soundness theorem:
\begin{lstlisting}
val run (e:ctree st a p q) : NMST a st.evolves (fun m -> st.interp p m) (fun _ x m' -> st.interp (q x) m')
\end{lstlisting}
where
\ls`NMST` is the effect of monotonic stateful
computations extended with nondeterminism. Here, we use it to
represent abstract, stateful computations whose states are constrained
to evolve according to the preorder \ls`st.evolves`, and which when
run in an initial state \ls`m` satisfying the interpretation of the
precondition \ls`p`, produce a result \ls`x` and final state \ls`m'`
satisfying the postcondition \ls`q x`. As such, using the Hoare types
of \ls`NMST`, the type of
\ls`run` validates the Hoare rules of CSL given by the indexing structure on
\ls`ctree`. In doing so, we avoid
the indirection of traces in \citepos{brookes04csl} original proof of
CSL as well as in the work of~\citet{nanevski14fcsl}.

\paragraph{Atomics and Invariants: Breaking circularities with monotonic state}
Although most widely used concurrent programming frameworks, e.g., the
POSIX pthread API, support dynamically allocated locks, few existing
CSL frameworks actually support them, with some notable
exceptions~\citep{buisse11storablelocks,gotsman07storable,hobor08oracle,JungKJBBD18,dodds16sync}.
The main challenge is to avoid circularities that arise from storing
locks that are associated with assertions about the memory in the
memory itself. Iris, with its step-indexed model of impredicativity,
can express this. However, other existing state of the art logics,
including FCSL, cannot. In \S\ref{sec:invariants}
and~\S\ref{sec:atomics}, we show how to leverage the underlying model of
monotonic state to allocate a stored invariant, and to open and close
it safely within an atomic command, without introducing
step indexing.

\paragraph{PCMs, ghost state, state machines, and implicit dynamic frames}
We base our memory model on partial commutative monoids (PCMs),
allowing the user to associate a PCM of their choosing with each allocation
unit.
Relying on \fstar's existing support for computationally irrelevant
erased types, we can easily model \emph{ghost state} by allocating
values of erased types in the heap, and manipulating these values only
using atomic ghost actions---all of which are erased during
compilation.
PCMs in \steelcore are orthogonal from ghost state: they can be used
both to separate and manage access permissions to both concrete and
ghost state---in practice, we use fractional permissions to control
read and write access to references.
Further, \steelcore includes a notion of \emph{monotonic} references,
which when coupled with \fstar's existing support for ghost values and
invariants, allow programmers to code up various forms of \emph{state
machines} to control the use and evolution of shared resources.
Demonstrating the flexibility of our semantics, we extend it to allow
augmenting CSL assertions with frameable heap predicates, a style that
combines CSL with \emph{implicit dynamic
frames}~\cite{smans12implicit} within the same mechanized framework.

\paragraph{Putting it to work} We present several examples
showing \steelcore at work, aiming to illustrate the flexibility and
extensibility of the logic and its smooth interaction with dependently
typed programming in \fstar. Starting with an atomic compare-and-set
(CAS) instruction, we program verified libraries for spin-locks, for
fork/join parallelism, and finally for protocol-indexed channel
types. Our channel-types library showcases dependent types at work
with \steelcore: its core construct is a type of channels,
\ls`chan p`, where \ls`p` is itself a free-monad-like
computation structure ``one-level up'' describing an infinite state
machine on types. We prove, once and for all, that programs using a
\ls`c:chan p` exchange a trace of messages on \ls`c` accepted by
the state machine \ls`p`.

\paragraph{Mechanization} \steelcore is a fully mechanized CSL
embedded in \fstar, and applicable to \fstar itself. The code and
proofs are available from \url{https://fstar-lang.org/papers/steelcore}.

\paragraph{Summary of contributions} In summary, the main
contributions of our work include the following:

\begin{itemize}
\item A new construction of indexed, effectful action trees,
mixing data and effectful computations to represent concurrent,
stateful and potentially divergent computations, with an indexing
structure capturing the proof rules of a generic CSL.

\item An intrinsically typed definitional interpreter that interprets
our effectful action trees into another effect, namely the effect of
nondeterminism layered on the effect of monotonic state. This provides
both a new style of soundness proof for CSL, as well as providing a
reference executable semantics for our host language \fstar extended
with concurrency.

\item An instantiation of our semantics with a modern CSL
inspired by recent logics like Iris, with a core memory model based on
partial commutative monoids and support for dynamically allocated
invariants. Relying on the underlying semantic model of monotonic
state is a key element, allowing us to internalize the step-indexing
that is necessary in Iris for dealing soundly with invariants.

\item We use our logic to build several verified libraries,
programmed in and usable by dependently typed, effectful host-language
programs, validating our goal of providing an Iris-style logic
applicable to a shallow rather than a deeply embedded programming
language.
\end{itemize}


\section{\fstar Background and Basic Indexed Action Trees}
\label{sec:background}

\fstar is a program verifier and a proof assistant based on
a dependent type theory with a hierarchy of
predicative universes (like Coq or Agda). \fstar also has a dependently typed
metaprogramming system inspired by Lean and Idris (called Meta-\fstar~\citep{metafstar})
that allows using \fstar itself to build and run tactics for constructing programs or proofs. More specific to \fstar is its effectful type
system, extensible with user-defined effects, and its use of SMT
solving to automate some proofs.

\paragraph*{Basic Syntax.} \fstar syntax is
roughly modeled on OCaml (\ls$val$, \ls$let$, \ls$match$ etc.)
although there are many differences to account for the additional
typing features.
Binding occurrences \ls$b$ of variables take the form \ls$x:t$, declaring
a variable \ls$x$ at type \ls$t$; or \ls$#x:t$ indicating that the
binding is for an implicit argument.
The syntax
\ls@fun (b$_1$) ... (b$_n$) -> t@ introduces a lambda abstraction, whereas
\ls@b$_1$ -> ... -> b$_n$ -> c@ is the shape of a curried function type.
Refinement types are written \ls$b{t}$,
e.g., \ls$x:int{x>=0}$ is the type of non-negative integers
(i.e., \ls$nat$).
As usual, a bound variable is in scope to the right of its binding; we
omit the type in a binding when it can be inferred; and for
non-dependent function types, we omit the variable name.
For example, the type of the
pure append function on vectors is written
\ls$#a:Type -> #m:nat -> #n:nat -> vec a m -> vec a n -> vec a (m + n)$,
with the two explicit arguments and the return type depending on the
three implicit arguments marked with `\ls$#$'. The type of pairs in \fstar is
represented by \ls`a & b`
with \ls`a` and \ls`b` as the types of the first
and second components respectively. In contrast, dependent tuple types are
written as \ls`x:a & b` where \ls`x`
is bound in \ls`b`. A dependent pair value is
written \ls`(| e, f |)` and we
use \ls`x._1` and \ls`x._2` for the first and second dependent
projection maps.

\subsection{A Total Semantics of Concurrency}
\label{sec:basic}

As an introduction to \fstar and a warm-up towards the main ideas
behind our indexed actions trees, we start by presenting a very simple
total semantics for concurrency.
Relying only on the pure rather than effectful features of \fstar,
some of the ideas in this section should also transfer to pure type
theories like Agda or Coq. However, our main construction involves a
partial-correctness semantics with effects like divergence, which may
be harder to develop in non-effectful type theories.

A disclaimer: total correctness for realistic concurrent programs
(e.g., under various scheduling policies) is a thorny issue that our
work does not address at all. For this introductory example, we focus
only on programs with structured parallelism, without any other
synchronization constructs, and where loop bounds do not depend on
effectful computations.

Our first step is to define a type of state-passing atomic
actions, \ls`action_tot a = state -> Tot (a & state)`. This is the
type of a function that transforms an initial \ls`state` to a pair of
an \ls`a`-typed result and a final \ls`state`.
%
The \ls`Tot` at the right of the arrow is a \emph{computation type}
emphasizing that this is a \emph{total} function; we will soon see
other kinds of computation types and effectful arrows in \fstar. All
unannotated arrows are \ls`Tot` by default.

\paragraph*{Action trees for concurrency} To model concurrency,
we define an inductive type \ls`ctree_total`, for trees
of \ls`action_tot` actions, indexed by a natural number (used for a
termination proof). This is our first and simplest instance of an
indexed action tree, one that could easily be represented in another
type theory. In \S\ref{sec:semantics}, we will enrich \ls`ctree_total`
to the CSL-indexed \ls`ctree` shown in \S\ref{sec:intro}.
\begin{lstlisting}
type ctree_total : nat -> Type -> Type =
| Ret  : #a:_ -> x:a -> ctree_total 0 a
| Act  : #a:_ -> act:action_tot a -> ctree_total 1 a
| Par  : (#aL #aR #nL #nR:_) -> ctree_total nL aL -> ctree_total nR aR -> ctree_total (nL+nR+1) (aL & aR)
| Bind : (#a #b #n1 #n2:_) -> f:ctree_total n1 a -> g:(x:a -> ctree_total n2 b) -> ctree_total (n1+n2+1) b
type nctree_total (a:Type) = n:nat & ctree_total n a
\end{lstlisting}

The type \ls`ctree_total` induces a monad-like structure (under a suitable
equivalence that quotients the use of \ls`Bind`) by representing
computations as trees of finite depth, with pure values (\ls`Ret`) and
atomic actions (\ls`Act`) at the leaves; a \ls`Bind` node for
sequential composition of two subtrees; and a \ls`Par` node for
combining a \ls`left` and a \ls`right` subtree. The monad induced
by \ls`ctree_total` differs from the usual construction of a free
monad for a collection of actions by including an explicit \ls`Bind`
node, instead of defining the monadic bind recursively. This
makes \ls`ctree_total` more similar
to \citepos{Pirog:ScopedOperations} scoped operations, with \ls$f$
being in the ``scope'' of \ls`Bind`. The \ls`nat` index counts the
number of \ls`Act`, \ls`Par` and \ls`Bind` nodes.
We
also define an abbreviation \ls`nctree_total a` to package a tree with
its index as a dependent pair.

\paragraph*{A definitional interpreter for \ls`ctree_total`}

To give a semantics to \ls`ctree_total`, we interpret its action trees
in an interleaving semantics for state-passing computations, relying
on a boolean tape to resolve the nondeterminism inherent in
the \ls`Par` nodes. To that end, we define a state and nondeterminism
monad, with \ls`sample`, \ls`get`, and \ls`put` actions:
\begin{lstlisting}
type tape = nat -> bool
type nst (a:Type) = tape & nat & state -> a & nat & state
let return (a:Type) (x:a) : nst a = fun (_, n, s) -> x, n, s
let bind (a b:Type) (f:nst a) (g:a -> nst b) : nst b = fun (t, n, s) -> let x, n1, s1 = f (t, n, s) in (g x) (t, n1, s1)
let sample () : nst bool = fun (t, n, s) -> t n, n+1, s
let get () : nst state = fun (_, n, s) -> s, n, s
let put (s:state) : nst unit = fun (_, n, _) -> (), n, s
\end{lstlisting}

We can now interpret \ls`ctree_total` trees as \ls`nst`
computations. It should be possible to define such an interpreter in
many type theories, in a variety of styles. Here, we show one way to
program it in \fstar, making use of its effect system to package
the \ls`nst` monad as a \emph{user-defined effect}.

A user-defined effect in \fstar introduces a new abstract computation
type backed by an existing \fstar definition (in our case, a
computation type \ls`NST` backed by the monad \ls`nst`). Based on work
by~\citet{swamy11coco}, computations and computation types enjoy some
conveniences in \fstar. In particular, \fstar automatically elaborates
sequencing and application of computations using the underlying
monadic combinators, without the need for do-notation, e.g., \ls`let`
in \ls`NST` is interpreted as \ls`bind` in \ls`nst`. Further, \fstar
supports sub-effects to lift between computation types, relying on a
user-provided monad morphism, e.g., pure computations are silently
lifted to any other effect.
The following incantation turns the \ls`nst` monad into the \ls`NST`
effect, with three actions, \ls`sample`, \ls`get` and \ls`put`.
\begin{lstlisting}
total new_effect { NST : a:Type -> Effect with repr=nst; return=return; bind=bind}
let sample () = NST?.reflect (sample()) $\quad$ let get () = NST?.reflect (get()) $\quad$ let put s = NST?.reflect (put s)
\end{lstlisting}
The type of \ls`sample` is \ls`unit -> NST bool`, where the
computation type at the right of the arrow indicates that
\ls`sample` has \ls`NST` effect---calling \ls`sample`
in a pure context is rejected by \fstar's effect system. We will soon
see examples of computation types with a richer indexing
structure. The \ls`total` qualifier on the first line ensures that all
the computations in the \ls`NST` effect are proved terminating.

Using \ls`NST`, we build an interpreter for \ls`ctree_total` trees by
defining \ls`run` as the transitive closure of a single \ls`step`.
The main point of interest is the last case of \ls`step`, reducing
a \ls`Par l r` node by sampling a boolean and recursing to evaluate a
step on either the left or the right.\ch{The Return? notation was never introduced}

\begin{lstlisting}
let reduct #a (r:nctree_total a) = r':nctree_total a{ Return? r' \/ r'._1 < r._1 }
let rec step #a (redex:nctree_total a) : NST (reduct redex) (decreases redex._1)
  match redex._2 with
  | Ret _ -> redex | Act act -> let s0 = get () in let x, s1 = act s0 in put s1; (| _, Ret x |)
  | Bind (Ret x) g -> (| _, g x |) | Bind f g -> let (| _, f' |) = step (| _, f |) in (| _, Bind f' g |)
  | Par (Ret x) (Ret y)  -> (| _, Ret (x, y) |)
  | Par l (Ret y) -> let (| _, l' |) = step (| _, l |) in (| _, Par l' (Ret y) |)
  | Par (Ret x) r -> let (| _, r' |) = step (| _, r |) in (| _, Par (Ret x) r' |)
  | Par l r ->
   if sample () then let (| _, l' |) = step (| _, l |) in (| _, Par l' r |) else let (| _, r' |) = step (| _, r |) in (| _, Par l r' |)
let rec run #a (p:nctree_total a) : NST (nctree_total a) (decreases p._1) = if Return? p then p else run (step p)
\end{lstlisting}

Having concluded our basic introduction to \fstar and indexed action
trees, we move beyond totality to general recursion and other effects,
and in \S\ref{sec:semantics} to indexed, effectful action trees.

\subsection{The Effects of Divergence and Monotonic State}
\label{sec:mst}

\paragraph{\ls`Dv` : an effect for divergence}
In addition to user-defined effects like \ls`NST`, \fstar
provides an abstract primitive effect of divergence represented by the
computation type \ls`Dv`. As with any other effect, the \ls`Dv` effect
is isolated from the logical core of \fstar: general recursive functions
in \ls`Dv` cannot mistakenly be used as proofs.~\citet{mumon} prove
the soundness of a core \fstar calculus in a partial correctness
setting for divergent computations, while also proving that \ls`Tot`
terms are normalizing. As such the following term is well-typed
in \fstar: \ls`let rec loop : unit -> Dv unit = fun () -> loop ()`.
From the perspective of \fstar's logical core, \ls`a -> Dv b` is an
abstract, un-eliminable type.

\paragraph{\ls`MST` : an effect for monotonic state} \ls`MST` is
another effect in \fstar for computations that read and write
primitive state, while restricting the state to evolve according to a
given preorder, i.e., a reflexive, transitive
relation.~\citet{preorders} observe that for such computations,
witnessing a property \ls`p` of the state that is invariant under the
preorder is sufficient to recall that \ls`p` is true in the
future. Ahman et al. propose the following signature for such
an \ls`MST` effect, and prove the partial correctness of the Hoare
logic encoded in the indexes of \ls`MST` against an operational
semantics for a $\lambda$-calculus with primitive state.

\begin{lstlisting}
effect MST (a:Type) (state:Type) (p:preorder state) (req:state -> prop) (ens:state -> a -> state -> prop)
\end{lstlisting}

When executing a computation \ls`(c : MST a state p req ens)` in an
initial state \ls`s0:state` satisfying
\ls`req s0`, the computation
either diverges, or returns a value \ls`x:a` in a final
state \ls`s1:state` satisfying
\ls`ens s0 x s1`. Further, the state is transformed
according to the preorder \ls`p`, i.e., the initial and final states
are related by \ls`p s0 s1`. The \ls`MST` effect provides the
following actions---for readability, we tag the pre- and postcondition
with \ls`requires` and \ls`ensures` respectively:

\vspace{1ex}
\noindent$\bullet~$\emph{Get} the current state:
\begin{lstlisting}
val get #state #p () : MST state state p (requires fun s -> True) (ensures fun s0 r s1 -> s0==s1 /\ r==s0)
\end{lstlisting}
\noindent$\bullet~$\emph{Put} the state, but only when the new state \ls`s1` is related to the old one \ls`s` by \ls`p`:
\begin{lstlisting}
val put #state #p (s1:state) : MST unit state p (requires fun s -> p s s1) (ensures fun _ _ s -> s==s1)
\end{lstlisting}
\noindent$\bullet~$\emph{Witness} stable predicates: A stable predicate is maintained across preorder-respecting state evolutions. The \ls`witness` action proves an abstract proposition,
\ls`witnessed q`, attesting that the stable predicate \ls`q` is valid.
\begin{lstlisting}
let stable_sprop #state (p:preorder state) = q:(state -> prop){forall s0 s1. q s0 /\ p s0 s1 ==> q s1}
val witnessed #state #p (q:stable_sprop p) : prop
val witness #state #p (q:stable_sprop p) : MST unit state p (fun s0 -> q s0) (fun s0 _ s1 -> witnessed q /\ s0==s1)
\end{lstlisting}

\noindent$\bullet~$\emph{Recall} stable predicates: Having \ls`witnessed q`, one can use \ls`recall q` to re-establish it at any point.
\begin{lstlisting}
val recall #state #p (q:stable_sprop p{witnessed q}) : MST unit state p (fun s0 -> True) (fun s0 _ s1 -> s0==s1 /\ q s1)
\end{lstlisting}

As such, the \ls`MST` effect provides a small program logic for
monotonic state computations, which we leverage for \steelcore's
semantic foundation in \S\ref{sec:steelcore}.

\paragraph{\ls`NMST`: extending \ls`MST` with nondeterminism}
The \ls`MST` effect only models state and does not provide the
nondeterminism we need for interleaving the subtrees of \ls`Par`
nodes. Therefore, we layer a user-defined effect of nondeterminism on
top of \ls`MST`, and define a new effect \ls`NMST` that provides an
additional \ls`sample` action---much as we did in the previous
section. We use \ls`NMST` in the next section as the target denotation
for the semantics of a generic partial correctness separation logic.

\paragraph{Erased types and the ghost effect} A final remark on \fstar's
effect system has to do with its support for erasure. As described
by~\citet{mumon}, \fstar encapsulates computations that are not meant
to be executed in a \emph{ghost effect}. Terms with ghost effect can
be used in proofs and specifications, but not in executable
code. Further, \fstar provides an extensible mechanism to mark certain
types as non-informative, including, notably the type
\ls`erased t`. Eliminating a term of a non-informative type (e.g., pattern
matching on it) incurs a ghost effect, ensuring that such uses never
occur in computationally relevant code. \fstar also supports an
implicit coercion mechanism that allows an \ls`erased t` to be used as
a \ls`t` (with a ghost effect)---such coercions are only legal in
computationally irrelevant contexts, e.g., proofs and
specifications. \fstar's extraction pipeline to several target
languages begins by erasing terms with non-informative types or ghost
effect to the unit value \ls`()`. In SteelCore, we rely on these
features implicitly. However, a full treatment of the erasure of
SteelCore terms for efficient extraction is beyond the scope of this
paper---indeed, we have yet to extract and run any Steel program,
though we do not foresee any major difficulties in doing so.

\section{Indexed Action Trees and a Partial Correctness Separation Logic}
\label{sec:semantics}

Recall from \S\ref{sec:intro} that our goal is to define the
indexed action trees with the following type:

\begin{lstlisting}
type ctree (st:state) (a:Type) (pre:st.slprop) (post:a -> st.slprop) : Type
\end{lstlisting}

The type is indexed by \ls`st:state`, a typeclass encapsulating
(at least) the type of the memory \ls`st.mem` and the type of
separation logic assertions on the memory \ls`st.slprop`. Intuitively,
a \ls`ctree st a fp0 fp1` is the type
of a potentially divergent, concurrent program manipulating shared
state of type \ls`st.mem`. The program expects the
\ls`fp0` footprint of some initial memory \ls`m0:st.mem`.
When run in \ls`m0`, it may diverge or produce
a \ls`result:a` and \ls`m1:st.mem`, providing the
\ls`(fp1 result)` fragment of \ls`m1` to the context.

The \ls`state` typeclass for the semantics is shown below. First, we
define a \ls`pre_state` containing all the operations we
need. A \ls`state` is a refinement of \ls`pre_state` satisfying
various laws.

\begin{lstlisting}
type pre_state = {  mem: Type; (* The type of the underlying memory *)
                 slprop: Type; (* The type of separation logic assertions *)
                 equals: equiv slprop; (* An equivalence relation on slprops *)
                 emp: slprop;  (* With a unit *)
                 star: slprop -> slprop -> slprop; (* And separating conjunction *)
                 interp: slprop -> mem -> prop; (* Interpreting slprop as a mem predicate *)
                 evolves: preorder mem; (* A preorder for MST: constrains how the state evolves *)
                 inv: mem -> slprop; (* A separation logic invariant on the memory *) }
let st_laws (st:pre_state) =
  associative st.equals st.star /\ commutative st.equals st.star /\ is_unit st.emp st.equals st.star /\
  interp_extensionality st.equals st.interp /\ star_extensionality st.equals st.star /\ affine st
type state = s:pre_state{st_laws s}
\end{lstlisting}

We expect \ls`emp` and \ls`star` to form a commutative monoid
over \ls`slprop` and the equivalence relation \ls`equals`. The
relation \ls`interp` interprets an \ls`slprop` as a predicate
on \ls`mem` and we expect the interpretation of \ls`star` to be
compatible with \ls`slprop`-equivalence.
We also expect the
interpretation to be \ls`affine`, in the sense that
\ls$interp (p `st.star` q) m ==> interp q m$.

As we will see in \S\ref{sec:memory} we can instantiate our semantics
with a separation
logic containing the full gamut of connectives, including conjunction,
disjunction, separating implication, and universal and existential
quantification. The preorder \ls`evolves` and the invariant \ls`inv`
are opaque as far as the semantics is concerned---in \S\ref{sec:steelcore},
we'll instantiate them in a way that allows us to model a source of
freshness for allocating ref cells and also to support dynamically allocated invariants.
In the following, we write \ls`*` for \ls`st.star` where \ls`st` is
clear from the context.

\subsection{Frame-Preserving Actions}
\label{sec:sem-action}

To define the type of action trees \ls`ctree`, let's start by defining
the type of atomic actions at the leaves of the tree:

\begin{lstlisting}
let action #st a (fp0:st.slprop) (fp1:a -> st.slprop) =
  unit -> MST a st.mem st.evolves
  (requires fun m0 -> st.interp (st.inv m0 * fp0) m0)
  (ensures fun m0 x m1 -> st.interp (st.inv m1 * fp1 x) m1 /\ preserves_frame fp0 (fp1 x) m0 m1)
\end{lstlisting}
An \ls`action` is an \ls`MST` computation that requires its initial
footprint \ls`fp0` to hold on the initial state \ls`m0`. It returns
an \ls`x:a` and ensures its final
footprint \ls`fp1 x` on the final state \ls`m1`. In both the pre- and
postcondition, we expect \ls`st.inv` to hold separately.
Finally, and
perhaps most importantly, the \ls`preserves_frame` side condition
ensures that actions are frameable. We elaborate on that next.

\paragraph{Frame preservation} We would like to derive a framing
principle for computations as a classic frame rule (and
its generalization, the rule for separating parallel composition). As
observed by~\citet{dinsdale-young13views}, it is sufficient for the
leaf actions to be frame-preserving for computations to be frame
preserving too. To that end, the definition of
\ls`preserves_frame` (that an action must provide in its postcondition)
states that all \ls`frame`s separate from \ls$st.inv m0 * pre$ and
valid in the initial state \ls`m0` remain separate from
\ls$st.inv m1 * post$ and are valid in \ls`m1`.
\begin{lstlisting}
let preserves_frame #st (pre post:st.slprop) (m0 m1:st.mem) =
  forall (frame:st.slprop). st.interp (st.inv m0 * pre * frame) m0 ==> st.interp (st.inv m1 * post * frame) m1
\end{lstlisting}

\subsection{CSL-Indexed Action Trees with Monotonic State}

\begin{figure}
\begin{lstlisting}
type ctree (st:state) : a:Type -> fp0:st.slprop -> fp1:(a -> st.slprop) -> Type =
| Act: e:action a fp0 fp1 -> ctree st a fp0 fp1
| Ret: fp:(a -> st.slprop) -> x:a -> ctree st a (fp x) fp
| Bind: f:ctree st a fp0 fp1 -> g:(x:a -> Dv (ctree st b (fp1 x) fp2)) -> ctree st b fp0 fp2
| Par: cL:ctree st aL fp0L fp1L -> cR:ctree st aR fp0R fp1R ->
     ctree st (aL & aR) (fp0L * fp0R) (fun (xL, xR) -> fp1L xL * fp1R xR)
| Frame: c:ctree st a fp0 fp1 -> f:st.slprop -> ctree st a (fp0 * f) (fun x -> fp1 x * f)
| Sub: c:ctree st a fp0 fp1 { sub_ok fp0 fp1 fp0' fp1' } -> ctree st a fp0' fp1'
$\text{with}$
let sub_ok #st fp0 fp1 fp0' fp1' = fp0' `stronger_than` fp0 /\ fp1' `weaker_than` fp1
let stronger_than #st fp0' fp0 = forall m f. st.interp (fp0' * f) m ==> st.interp (fp0 * f) m
let weaker_than #st fp1' fp1 = forall x m f. st.interp (fp1 x * f) m ==> st.interp (fp1' x * f) m
\end{lstlisting}
\caption{\steelcore's representation of computations as indexed action trees}
\label{fig:m}
\end{figure}

Figure~\ref{fig:m} shows the way we represent computation trees in
\steelcore---extending the \ls`ctree` type from the introduction.
To reduce clutter, we omit binders for implicit
arguments---in the type of each constructor of the inductive \ls`ctree`,
we only bind names that do not appear free in other arguments of the
constructor.
These trees differ from the simple action trees we used in
\S\ref{sec:basic}.
The additional
indexing structure in each case of \ls`ctree` posits the proof rules
of a program logic for reasoning about \ls`ctree`
computations. In \S\ref{sec:run}, we show that this logic is
sound by denoting \ls`ctree st a fp0 fp1` trees via an interleaving,
definitional interpreter into \ls$NMST$ computations.
As \ls$NMST$ computations are
potentially divergent, we do not need to prove termination of the definitional
interpreter. Thus the type \ls$ctree$ does not
carry a natural number index as we did in \S\ref{sec:basic}.

We describe the structure of \ls`ctree` in detail, discussing each of its
constructors in turn.

\paragraph{Atomic actions} At the leaves of the tree,
we have nodes of the form \ls`Act e`, for some action \ls`e`: the
index of the computation inherits the indexes of the action.

\paragraph{Returning pure values} Also at the leaves
of the tree are \ls`Ret fp x` nodes, which allow returning a pure
value \ls`x` in a computation. The \ls`Ret` node is parametric in a
footprint \ls`fp`, and the indexes on \ls`ctree`
state that in order to provide \ls`fp`, we expect the
\ls`fp x` to hold in the initial state \ls`m0`. An
alternative formulation could also have used
\ls`Ret : x:a -> ctree st a st.emp (fun _ -> st.emp)`,
although, as we discuss in \S\ref{sec:run}, this form
is less convenient in conjunction with the frame rule.

\paragraph{Sequential composition} The \ls`Bind f g` node sequentially
composes \ls`f` and \ls`g`. Its indexing structure should appear
fairly canonical. The footprints of \ls`f` and \ls`g` are ``chained''
as in~\citepos{atkey09parameterised} parameterized monads, except our
indexes (notably \ls`fp2`) are dependent. The computation type
of \ls`g` has the \ls`Dv` effect, indicating a potentially divergent
continuation.

\paragraph{Parallel composition} \ls`Par cL cR` composes
computations in parallel. The indexing structure yields the classic
CSL rule for parallel composition of computations with disjoint
footprints.

\paragraph{Structural rules: Framing and Subsumption}
The \ls`Frame c f` node preserves the frame \ls`f` across the computation
\ls`c`. The \ls`Sub c` node allows
strengthening the initial footprint and
weakening the final footprint of \ls`c`. These nodes
directly correspond to the canonical CSL frame and consequence rules.

These structural rules are essential elements of our
representation. The indexing structure of \ls`ctree` defines a program
logic and the structural rules are manifested as a kind of re-indexing,
which must be made explicit in the inductive type as additional
constructors. Further, given such structural rules, the need for a
separate \ls`Bind`, as opposed to continuations in each node, becomes
evident.
Consider verifying a Hoare triple
\ls`{P1 * P} a1; a2; a3 {Q}`,
where \ls`a1`, \ls`a2`, \ls`a3` are actions with
\ls`{ P1 } a1; a2 { P1 }`, and
\ls`{ P1 * P } a3 { Q }`. The canonical proof frames \ls`P` across
\ls`a1; a2` together, which is trivial to do with our representation,
as
\ls`Bind (Frame (Bind (Act a1) (fun _ -> (Act a2))) P) (fun _ -> (Act a3))`.
The frames can be easily added \emph{outside} of a
proof derivation, making the proofs modular.
However, if the continuations were part of the \ls`Act` (and \ls`Par`)
nodes,
such a structural frame rule would not apply. We would have to bake-in
framing in the \ls`Act` nodes, and even then we would have to frame \ls`P`
across \ls`a1` and \ls`a2` individually. This makes the proofs less
modular, since we can't directly use the given derivation
\ls`{ P1 } a1; a2 { P1 }`.

Although we include \ls$Frame$ and \ls$Sub$, we lack the structural
rule for disjunction. Accommodating disjunction in a shallow embedding
is hard to do, since it requires giving to the same computation more than
one type. One possibility may be to adopt a relational specification
style, as~\citet{NanevskiVB10} do---we leave an exploration of this
possibility to future work. Meanwhile, as we instantiate the semantics
with a state model in \S\ref{sec:steelcore}, we also provide several
lemmas to destruct combinations of separating conjunctions and
existentials (with disjunctions as a special case).\ch{This last phrase
  not clear to me.}

\subsection{Soundness}
\label{sec:run}

To prove the soundness of the proof rules induced by the indexing
structure of \ls`ctree`, we follow the strategy outlined
in \S\ref{sec:basic}, with \ls`NMST` from \S\ref{sec:mst} as the
target denotation. Our goal is to define an interpreter with the
following type, showing that it maintains the memory invariant while
transforming \ls`fp0` to \ls`fp1 x`.
\begin{lstlisting}
val run #st #a #fp0 #fp1 (f:ctree st a fp0 fp1) : NMST a st.mem st.evolves
  (requires fun m0 -> st.interp (st.inv m0 * fp0) m0) (ensures fun m0 x m1 -> st.interp (st.inv m1 * fp1 x) m1)
\end{lstlisting}

As before, we proceed by first defining a single-step interpreter and
then closing it transitively to build a general recursive, multi-step
interpreter. The single-step interpreter has the following type, returning
(as in~\S\ref{sec:basic}) the reduced
computation tree packaged with all its indices.
\begin{lstlisting}
type reduct #st a = | Reduct: fp0:_ -> fp1:_ -> ctree st a fp0 fp1 -> reduct a
val step (f:ctree st a fp0 fp1) : NMST (reduct a) st.mem st.evolves
  (requires fun m0 -> st.interp (st.inv m0 * fp0) m0)
  (ensures fun m0 (Reduct fp0' fp1' _) m1 ->
    st.interp (st.inv m1 * fp0') m1 /\ preserves_frame fp0 fp0' m0 m1 /\ fp1' `stronger_than` fp1)
\end{lstlisting}
In addition to requiring and ensuring the invariant and footprint
assertions, we have additional inductive invariants that are needed to
take multiple steps. As is typical in such proofs, one needs to show
that given a term in a context \ls`E[c]`, reducing \ls`c` by a single
step produces \ls`c'` that can be correctly typed within the same
context, i.e., \ls`E[c']` must be well-typed. Towards that end, we
need two main properties of \ls`step`: (a) \ls`preserves_frame`,
defined in \S\ref{sec:sem-action}, ensures that the reduct \ls`c'` can
be framed with any frame used with the redex \ls`c`; and (b) that the
postcondition \ls`fp1'` of the reduct \ls`c'` is stronger than the
postcondition \ls`fp1` of the redex \ls`c`. Interestingly, we don't
explicitly need to show that the precondition of the reduct is weaker
than the precondition of the redex: that the initial footprint of the
reduct holds in \ls`m1` is enough.
We show all the main cases of
the single-step reduction next. In all cases, the code is typechecked
as shown, with proofs semi-automated by \fstar's SMT solving backend.

\paragraph{Framing} The code below shows stepping
through applications of the \ls`Frame c0 f` rule. In the case
where \ls`c0` is a \ls`Ret` node, we remove the \ls`Frame` node and
restore the derivation by extending the footprint of the \ls`Ret` node
to include the frame \ls`f`---this is one reason why it is convenient
to have \ls`Ret` nodes with parametric footprints, rather than just
the \ls`emp` footprint.
\begin{lstlisting}
let rec step #st #a #fp0 #fp1 (c:ctree st a fp0 fp1) = match c with  | $\ldots$
  | Frame (Ret fp0' x) f -> Reduct (fp0' x * f) (fun x -> fp0' x * f) (Ret (fun x -> fp0' x * f) x)
  | Frame c0 f -> let m0 = get () in let Reduct fp0' fp1' c' = step c0 in let m1 = get () in
                preserves_frame_star fp0 fp0' m0 m1 f; Reduct (fp0' * f) (fun x -> fp1' x * f) (Frame c' f)
\end{lstlisting}
When \ls`c0` is not a \ls`Ret`, we recursively evaluate
a step within \ls`c0` and then reconstruct a \ls`Frame` around its
reduct \ls`c'`. This proof step makes use of a key
lemma, \ls`preserves_frame_star`, which states
\ls$preserves_frame fp0 fp0' m0 m1 ==> preserves_frame (fp0 * f) (fp0' * f) m0 m1)$.

\paragraph{Subsumption} Reductions of the other structural rule,
\ls`Sub`, is simpler, we just remove the \ls`Sub` node, as shown
below; the refinement \ls`sub_ok` on the \ls`c` argument of
the \ls`Sub` node allows \fstar to prove the inductive invariants
of \ls`step`. Although we remove \ls`Sub` nodes, the rule for
sequential composition (next) adds them back to ensure that the reduct
remains typeable in context. An alternative may have been to
treat \ls`Sub` like we treat \ls`Frame`, however, this form is more
convenient when adding support for implicit dynamic frames (as mentioned briefly below).
\begin{lstlisting}
  | Sub #fp0' #fp1' c -> Reduct fp0' fp1' c
\end{lstlisting}

\paragraph{Sequential composition} In case \ls`f` is fully reduced to
a \ls`Ret` node, we simply apply the
continuation \ls`g`. Otherwise, we take a step in \ls`f` producing a
reduct \ls`f'` that may have a stronger final footprint. To reconstruct
the \ls`Bind` node, we need to strengthen the initial footprint of \ls`g`
with the final footprint of \ls`f'`, we do so by wrapping \ls`g` with a \ls`Sub`:
\begin{lstlisting}
  | Bind #fp2 (Ret fp0 x) g -> Reduct (fp0 x) fp2 (g x)
  | Bind #fp0 #fp1 #fp2 f g -> let Reduct fp0' fp1' f' = step f in
                           Reduct fp0' fp2 (Bind f' (Sub #fp1 #_ #fp1' #_ g))
\end{lstlisting}

\paragraph{Parallel composition} The structure of reducing \ls`Par` nodes
is essentially the same as in~\S\ref{sec:basic}. When both
branches are \ls`Ret` nodes, we simply create a reduct with a \ls`Ret`
node capturing the two values.

\begin{lstlisting}
  | Par (Ret fp0L xL) (Ret fp0R xR) ->
    Reduct (fp0L xL * fp0R xR) (fun (xL, xR) -> fp0L xL * fp0R xR) (Ret (fun (xL, xR) -> fp0L xL * fp0R xR) (xL, xR))
  | Par #aL #fp0L #fp1L cL #aR #fp0R #fp1R cR ->
    if sample() then let m0 = get () in let Reduct fp0L' fp1L' cL' = step cL in let m1 = get () in
                   preserves_frame_star fp0L fp0L' m0 m1 fp0R;
                   Reduct (fp0L' * fp0R) (fun (xL, xR) -> fp1L' xL * fp1R xR) (Par cL' cR)
    else ...  (* similarly for the right branch *)
\end{lstlisting}

When only one of the branches is \ls`Ret`, we descend into
the other one (we elide these cases from the presentation). When
both the branches are candidates for reduction,
we sample a boolean and pick either the left or right branch to
descend into. Having obtained a reduct, we reconstruct the \ls`Par`
node, by appropriately framing the initial footprint of the unreduced
branch, as shown above. 

\paragraph{Atomic actions} An \ls`Act e` node is reduced by
applying it, and returning its result in a \ls`Ret` node.

\begin{lstlisting}
  | Act #fp1 e -> let x = e () in Reduct (fp1 x) fp1 (Ret fp1 x)
\end{lstlisting}

\paragraph{Multi-step interpreter}
Implementing a general recursive, multi-step interpreter is
straightforward: we recursively evaluate single steps until we reach
a \ls`Ret` node. The type of the interpreter, shown below, is the main
statement of partial correctness for our program logic.
\begin{lstlisting}
let rec run #st #a #fp0 #fp1 (f:ctree st a fp0 fp1) : NMST a st.mem st.evolves
  (requires fun m0 -> st.interp (st.inv m0 * fp0) m0) (ensures fun m0 x m1 -> st.interp (st.inv m1 * fp1 x) m1)
= match f with | Ret _ x -> x | _ -> let Reduct _ _ f' = step f in run f'
\end{lstlisting}
The type states that when run in an initial state \ls`m0` satisfying
the memory invariant \ls`st.inv m0` and separately the
footprint assertion \ls`fp0`,
the code either diverges or returns \ls`x:a` in a final state \ls`m1`
with the invariant \ls`st.inv m1`, and the footprint assertion
\ls`fp1 x`.
The inductive \ls`stronger_than` invariant about the \ls`step`
function providing a stronger postcondition is crucial to the proof
here: the recursive call to \ls`run` ensures the validity of the
post-footprint of \ls`f'` in the final memory, we need the
inductive invariant to relate it to the post-footprint
of \ls`f`, as required\ch{ensured?} by the postcondition of \ls`run`.

\paragraph{Extension: Implicit Dynamic Frames}
While we have presented the action trees, and hence the CSL
semantics, using only the \ls`slprop` indices, our actual
implementation also contains two further indices for specifications in
the style of implicit dynamic frames~\cite{smans12implicit}. In our
representation type, \ls`ctree_idf st a fp0 fp1 req ens`, the last two
indexes \ls`req` and \ls`ens` indicate the pre- and postcondition of a
computation, where the precondition is a \emph{\ls`fp0`-dependent}
predicate on the initial memory and postcondition is a two-state
predicate that is \ls`fp0`-dependent on the initial memory
and \ls`fp1`-dependent on the final one.  The dependency relation
captures the requirement that the predicates are
``self-framing''~\citep{ParkinsonSummers12}, i.e., the \ls`slprop`
footprint indices \ls`fp0` and \ls`fp1` limit the parts of the memory
that these predicates can depend on.
In addition to these indices, we add frameable memory predicates to
the \ls`Frame` and \ls`Par` rule.

For example, the following is
the \ls`Frame` rule in our implementation (\ls`fp_prop f` is the type
of an \ls`f` dependent memory predicate):

\begin{lstlisting}
  | Frame: c:ctree_idf st a fp0 fp1 req ens -> f:st.slprop -> p:fp_prop f ->
          ctree_idf st a (fp0 * f) (fun x -> fp1 x * f) (frame_req req p) (frame_ens ens p)
$\text{with}$
let frame_req req p = fun m -> req m /\ p m  $\quad\quad\quad\quad$ let frame_ens ens p = fun m0 x m1 -> ens m0 x m1 /\ p m1
\end{lstlisting}

Finally, our soundness theorem, i.e. the specification of the \ls`run`
function, requires the precondition (\ls`req`) of the computation and
ensures its postcondition (\ls`ens`). To prove this theorem, we have
to enhance the inductive invariant of the \ls`step` function to also
mention weakening of the preconditions and strengthening of the
postconditions. By incorporating both the \ls`slprop` indices, and
implicit dynamic frame-style requires- and ensures-indices, we hope to
provide more flexibility in writing specifications for client
programs.

\paragraph{Discussion}
It's worth noting that although we've built a definitional interpreter
with an interleaving semantics for concurrent programs, we do not
intend to run programs using \ls`run`, since it would be very
inefficient, primarily because the interleaving semantics is encoded
via sampling, but also because the representation includes a full
proof tree, including the structural rules like \ls`Frame`
and \ls`Sub`. Instead, relying on \fstar's support for extraction to
OCaml and C, we intend to compile effectful, concurrent programs to
native concurrency in the target platforms, e.g., POSIX threads. As
such, the main value of \ls`run` is its proof of soundnesss: we now
have in hand a semantics for concurrent programs and a means to reason
about them deductively using a concurrent separation logic. We've
built our semantics atop the effect of monotonic state, parameterizing
our semantics with a preorder governing how the state evolves.  So
far, this preorder has not played much of a role. For the payoff,
we'll have to wait until we instantiate the \ls`state` interface,
next.

\section{The SteelCore Program Logic}
\label{sec:steelcore}

The core semantics developed in the previous section provides a
soundness proof for a generic, minimalistic concurrent separation
logic. In this section, we instantiate the semantics with a model of
state, assertions, invariants and actions defining the logic
for \steel programs.

The logic includes the following main features:

\begin{itemize}
\item A core heap model addressed by typed references with
      explicit, manually managed lifetimes.

\item Each heap cell stores a value in a user-chosen, cell-specific
      partial commutative monoid, supporting various forms of sharing
      disciplines and stateful invariants, including, e.g., a
      discipline of fractional permissions~\citep{boyland03frac}, for
      sharing among multiple threads.

\item A separation logic, with all the usual connectives.

\item Ghost state and ghost actions,
      relying on \fstar's existing support for erasure.

\item A model of atomic actions, including safe
      composition of ghost and concrete actions.

\item Invariants, that can be dynamically allocated and
      freely shared among multiple threads and accessed and restored
      by atomic actions only.

\item Monotonic references controlling the evolution of memory,
      built using preorders from the underlying monotonic state
      effect.
\end{itemize}

The result is a full-featured separation logic shallowly embedded
in \fstar, with a fully mechanized soundness proof, and applicable
directly to dependently typed, higher order, effectful host language
programs. We provide several small examples throughout this section,
and further ones in \S\ref{sec:examples}.

\subsection{Memory}
\label{sec:memory}

At the heart of our state model is a representation of memory, as
outlined in the type below.

\begin{lstlisting}
let addr = nat $\qquad$ let heap = addr -> option cell $\qquad$ type mem = {  heap:heap; ctr:nat; istore:istore }
let mem_inv m : slprop = (forall i.i >= m.ctr ==> m.heap i == None) /\ istore_inv m
let mem_evolves m$_0$ m$_1$ = h_evolves m$_0$.heap m$_1$.heap /\ i_evolves m$_0$.istore m$_1$.istore /\ m$_0$.ctr <= m$_1$.ctr
\end{lstlisting}

A \ls`heap` is a map from abstract addresses (\ls`nat`) to
heap \ls`cell`s defined below.\footnote{In our \fstar sources, we
define \ls`heap` as the type \ls`addr ^-> option cell`, the type of
functions for which the functional extensionality axiom is admissible
in \fstar; we gloss over this technicality in our presentation here.}
A memory augments a heap with two important fields of metadata. First,
we have a counter to provide fresh addresses for allocation (with an
invariant guaranteeing that all addresses above \ls`ctr` are
unused). Second, we have an \ls`istore` for tracking dynamically
allocated invariants. Actions maintain a memory invariant \ls`inv` and
the memory is constrained to evolve according to the
preorder \ls`mem_evolves`. We discuss all these elements in detail
throughout this section.

\newcommand\pleq[1][p]{\ensuremath{\preceq_{\mathsf{#1}}}}

For the definition of heap cells, we make use of partial
commutative monoids (PCMs).
Using PCMs to represent state is typical in the literature:
starting at least with the work of~\citet{jensen12fictional}, PCMs have
been used to encode a rich variety of specifications, ranging from
various kinds of sharing disciplines, fictional separation, and also
various forms for state machines.
We represent PCMs as the typeclass \ls`pcm a` shown below, where we account
for partiality by restricting the domain of \ls`op` by a
predicate \ls`composable`. We write $\pleq$ for the partial
order induced by \ls`p:pcm a`.

\begin{lstlisting}
type pcm (a:Type) = { one:a; composable:  a -> a -> prop {sym composable};
                    op: x:a -> y:a{composable x y} -> a { comm op /\ assoc op /\ is_unit op one } }
let ($\preceq$) (#a:Type) (#pcm:pcm a) (x y : a) = exists frame. pcm.op x frame == y
type cell = | Cell: a:Type -> pcm:pcm a -> v:a -> cell
\end{lstlisting}

A cell is a triple of a type \ls`a`, an instance of the
typeclass of partial commutative monoids (\ls`pcm a`), and a value of
that type.%
\footnote{On universes and higher order stores: We define our memory model
universe-polymorphically, so that it can store values in higher
universes, e.g., values at existential
types (\S\ref{sec:channels}). However, the \ls`cell` type resides in a universe one greater
than the type it contains. By extension, \ls`heap` is in the same
universe as its cells. As a result, although \ls`heap`s
and \ls`heap`-manipulating total functions cannot be stored in cells,
functions in \fstar that include the effect of divergence are always
in universe \ls`0` and can be stored in the heap, i.e., this model is
adequate for partial correctness of programs with higher order
stores.}

With this representation of \ls`heap`, it's relatively straightforward
to define two functions, \ls`disjoint` and \ls`join`, which we use to
separate and combine disjoint memories.\ch{Obscure phrasing: does the which
  apply to both of them or just to join? Are you running join in reverse
  to separate memories?}
For an address that appears in both heaps, we require the cell
at that address to agree on the type, the PCM instance, and for the
values to be composable in the PCM.

\begin{lstlisting}
let disjoint_addr (h h':heap) (a:addr) = match h a, h' a with
  | Some (Cell t pcm v), Some (Cell t' pcm' v') -> t==t' /\ pcm==pcm' /\ pcm.composable v v' | _ -> True
let disjoint h0 h1 = forall a. disjoint_addr h0 h1 a
let join (h0:heap) (h1:heap{disjoint h0 h1}) = fun a -> match h0 a, h1 a with
  | None, None -> None | None, Some x | Some x, None -> Some x
  | Some (Cell t pcm v0), Some (Cell _ _  v1) -> Some (Cell t pcm (pcm.op v0 v1))
\end{lstlisting}

\subsection{Separation Logic Propositions}
\label{sec:slprop}

\nik{New text here}

We define the type \ls`slprop` of separation logic propositions as the
type of heap propositions \ls`p` that are preserved under disjoint
extension. We emphasize that \ls`slprop`s are affine \emph{heap}
propositions, rather than \ls`mem` propositions---the non-heap fields
in a memory (e.g., freshness counters etc.) are meant for internal
bookkeeping and (intentionally) cannot be described by slprops. We
use \ls`interp` to apply an \ls`slprop` to the heap within a
memory. Further, being heap predicates, \ls`slprop`s reside in the
same universe as \ls`heap`. As such, \ls`slprops` cannot themselves be
stored in the heap, although doing so is sometimes convenient for
encoding various forms of higher-order ghost
state~\citep{jung16higherorder}---this is the main limitation of our
model. However, since Steel is embedded within \fstar, one can
sometimes work around this restriction by adopting various dependently
typed programming tricks, e.g., rather than storing \ls`slprops` in
the heap, one might instead store codes for a suitably small
sub-language of \ls`slprops` instead and work with interpretations of
those codes.

\begin{lstlisting}
let slprop = p:(heap -> prop) { forall (h0 h1: heap). p h0 /\ disjoint h0 h1 ==> p (join h0 h1) }
let interp (p:slprop) (m:mem) = p m.heap
\end{lstlisting}

We define several basic connectives for \ls`slprop`, as shown
below. The existential and universal quantifiers, \ls`slex`
and \ls`slall` support quantification over terms in arbitrary
universes, including quantification over \ls`slprop`s themselves.

\begin{lstlisting}
let slstar p1 p2 h = exists h1 h2.  h1 `disjoint` h2 /\ h == join h1 h2 /\ p1 h1 /\ p2 h2
let slwand p1 p2 h = forall h1. h `disjoint` h1 /\ p1 h1 ==> p2 (join h h1)
let slemp p h = True $\quad\quad\quad\,$ let sland p1 p2 h = p1 h /\ p2 h $\quad$ let slor p1 p2 h = p1 h \/ p2 h
let slex p h = exists x. p x h $\quad$ let slall p h = forall x. p x h
\end{lstlisting}

This interpretation also induces a natural equivalence relation
on \ls`slprop`, i.e.,
\ls`p $\sim$ q` iff \ls`(forall m. interp p m` \ls`<==> interp q m)`
 (extensional equivalence of heap predicates) and it is easy to prove
that \ls`star` and \ls`emp` form a (total) commutative monoid with
respect to $\sim$.

We also define the atomic points-to assertion on references.
\begin{lstlisting}
let ref (a:Type) (p:pcm a) = addr
let pts_to (r:ref a p) (v:a) (h:heap) = match h r with Some (Ref a' p' v') -> a == a' /\ p == p' /\ v $\preceq_p$ v' | _ -> False
val pts_to_compatible (r:ref a p) (v0 v1:a) (m:mem) : Lemma
    (interp (pts_to r v0 `star` pts_to r v1) m <==> (p.composable v0 v1 /\ interp (pts_to x (p.op v0 v1)) m))
\end{lstlisting}
A reference is represented by its address in the heap and
\ls`pts_to r v` asserts partial knowledge of the contents of
the reference \ls`r`, i.e., that
\ls`r` contains some value \ls`v'` compatible with \ls`v` according
to the PCM associated with \ls`r`.  The \ls`pts_to_compatible` lemma
relates the separating conjunction to composition in the underlying
PCM. In coming sections we will see how to choose specific PCMs to
model fractional permissions and monotonic references.

We now have most of what we need to instantiate the \ls`state`
interface of our semantics---two key ingredients, the memory invariant
and preorder will be presented in detail in the the next three
subsections. Foreshadowing their presentation, our state instantiation is:

\begin{lstlisting}
let st : state = { mem = mem; slprop = slprop; equals = $\sim$; emp = slemp; star = slstar; interp = interp;
              inv = mem_inv (* cf. $\S\ref{sec:invariants}$ *); evolves = mem_evolves (* cf. $\S\ref{sec:atomics},\ref{sec:monotonic-refs}$ *) }
\end{lstlisting}

\paragraph{Actions on PCM-indexed references} Given this instantiation,
one can define several basic actions (cf. \S\ref{sec:sem-action}), such as the following primitives
on references. Building on these generic primitives, we implement
libraries for several more common use cases, including references with
fractional permissions and monotonic references.

To allocate a reference, one presents both a value and a PCM to use
for that reference.
\begin{lstlisting}
val alloc (#p:pcm a) (v:a) : action (ref a p) emp (fun r -> pts_to r v)
\end{lstlisting}
Reading a reference with \ls`(!)` returns a value compatible with the
caller's partial knowledge.
\begin{lstlisting}
val (!) (r:ref a p) (v:erased a) : action (x:a{v $\preceq_p$ x}) (pts_to r v) (fun v' -> pts_to r v)
\end{lstlisting}
Mutating a reference \ls`r` requires the new value \ls`v` to be
compatible with all frames compatible with the caller's partial
knowledge of \ls`r`.
\begin{lstlisting}
let frame_preserving (#p:pcm a) (x y:a) = forall f. p.composable f x ==> p.composable f y /\ p.op f y == y
val (:=) (r:ref a p) (v0:erased a) (v:a{frame_preserving v0 v}) : action unit (pts_to r v0) (fun _ -> pts_to r v) $\label{line:frame-preserving-update}$
\end{lstlisting}
Finally, to de-allocate a reference the caller must possess exclusive non-trivial knowledge of it.
\begin{lstlisting}
let exclusive (#p:pcm a) (v:erased a) = forall frame. p.composable frame v0 ==> frame==p.one
val free (r:ref a p) (v0:erased a{exclusive v0}): action unit (pts_to r v) (fun _ -> emp)
\end{lstlisting}

In what follows, we overload the use of \fstar's existing
connectives \ls`exists, forall, /\, \/` for use with \ls`slprop`.
We write \ls`emp` for \ls`slemp`; \ls$`star`$ and $\wand$
for \ls`slstar` and \ls`slwand`.
Borrowing \fstar's notation for refinement types, we also
write \ls`h:p{f}` for \ls`sland p (fun h -> f)` and \ls`pure p`
for \ls`_:emp{p}`.

\subsection{Introducing Invariants: Preorders and the \ls`istore`}
\label{sec:invariants}

Beyond the traditional separation logic assertions, it is useful to
also support a notion of \emph{invariant} that allows a
non-duplicable \ls`slprop` to be
shared among multiple threads. For some basic intuition, it's
instructive to look at the design of invariants in Iris---we
reproduce, below, three of~\citepos{JungKJBBD18} rules related to
invariants (slightly simplified).
\newcommand\IrisInvName{\ensuremath{\mathcal{N}}}
\newcommand\IrisInvSet{\ensuremath{\mathcal{E}}}
\newcommand\IrisUpdate[1][\IrisInvSet]{\ensuremath{\Rrightarrow}_{#1}}
\newcommand\IrisInv[2][\IrisInvName]{\ensuremath{\mbox{\fbox{$#2$}}^{#1}}}
\newcommand\IrisTriple[4][\IrisInvSet]{\left\{~#2~\right\}~#3~\left\{~#4~\right\}_{#1}}
\newcommand\IrisNext{\ensuremath\triangleright}
\arraycolsep=4pt
\[\begin{array}{ccc}
(1)~P \IrisUpdate \IrisInv{P} &
(2)~\mbox{persistent}\left(\IrisInv{P}\right) \\
~\\
    \multicolumn{2}{c}{
(3)~\inference{\IrisTriple[\IrisInvSet \setminus \IrisInvName]{\IrisNext P \ast Q}{e}{\IrisNext P \ast R} &
           \mbox{atomic(}e\mbox{)} &
           \IrisInvName \subseteq  \IrisInvSet}
          {\IrisTriple{\IrisInv{P} \ast Q}{e}{\IrisInv{P}\ast R}}} \\
\end{array}\]

The first rule states that at any point, one can turn a resource
assertion $P$ into an \emph{invariant} $\IrisInv{P}$. An invariant is
associated with a name, \IrisInvName---we shall see its significance
in \S\ref{sec:atomics}.

The second rule states that an invariant is persistent, which implies it is duplicable:
i.e., $\IrisInv{P} \implies \IrisInv{P} \ast \IrisInv{P}$. Thus, by
turning a resource assertion $P$ into an invariant, one can share the
invariant among multiple threads, frame it across other computations,
etc.

The final rule shows how an invariant can be used. This rule is quite
technical, but intuitively it states that an atomic
command $e$ can assume the resource assertion $P$ associated with an
invariant $\IrisInv{P}$, so long as it also restores $P$ after
executing (atomically). Some of the technicality in the rule has to do
with impredicativity and step indexing. In Iris, $\IrisInv{P}$ is a
proposition in the logic like any other, and Iris allows
quantification over all such propositions, including invariants
themselves. This is very powerful, but it also necessitates the use of
step indexing, i.e., the ``later'' modality $\IrisNext{P}$ in the premise of the
rule. For \steelcore, we seek to model invariants of a similar flavor,
but while remaining in our predicative setting---our use of the
monotonic state effect will give us a way.

\paragraph{Invariants in \steelcore} To allocate an invariant,
we provide an action with the signature below:
\begin{lstlisting}
val new_invariant (p:slprop) : action (ival p) p (fun _ -> emp)
\end{lstlisting}
Recall the \ls`action` type from \S\ref{sec:sem-action}. The type
above states that given possession of \ls`p`, \ls`new_invariant`
consumes \ls`p`, providing only \ls`emp`, but importantly, returning a
value of type \ls`ival p`: our representation of an
invariant---\ls`new_invariant` models Iris' update modality to
allocate an invariant, i.e., the first of the three rules above.
Being a value, \ls`ival p` is freely duplicable, like any other value
in \fstar---mimicking Iris' rule of persistence of invariants.
Finally, sketching (imprecisely) what we develop in detail
in \S\ref{sec:atomics}, we provide a combinator below that is the
analog of Iris' rule for eliminating and restoring invariants in
atomic commands---an atomic command that expects \ls$p `star` q$ and provides
\ls$p `star` r$ can be turned into a command that only expects \ls`q` and
provides \ls`r`, as long as an \ls`ival p` value can be presented as
evidence that \ls`p` is an invariant.\ch{This type is different (much simpler)
  from the one in 4.4}
\begin{lstlisting}
val with_invariant (i:ival p) (e:atomic a (p `star` q) (p `star` r)) : atomic a q r
\end{lstlisting}

\paragraph{Representing Invariants} We will use
the \ls`istore` component of a \ls`mem` to keep track of invariants
allocated with \ls`new_invariant`: an \ls`istore` is a list
of \ls`slprop`s and the name associated with an invariant is its
position in the list.
The invariant of
the \ls`istore` (included in \ls`inv`, which, recall
from \S\ref{sec:semantics}, is expected and preserved by every step of
the semantics) requires every invariant in the \ls`istore` to be
satisfied separately.
The \ls`i_evolves` preorder (part of the \ls`mem_evolves` preorder
shown in \S\ref{sec:memory}) states that when the memory evolves,
the \ls`istore` only grows. The predicate \ls`inv_for_p i p m` states
that the invariant name \ls`i` is associated with \ls`p` in the
memory \ls`m`---its stable form, \ls`i~>p`, makes use of
the \ls`witnessed` connective used with the \ls`MST` effect introduced
in \S\ref{sec:mst}. \ls`i~>p` is the \steelcore equivalent of
$\IrisInv[i]{p}$, i.e., the name \ls`i` is always associated with
invariant \ls`p`. Since \ls`i~>p` is just a \ls`prop`, it is naturally
duplicable. It's also convenient to treat invariants as a value
type, \ls`ival p`---just an invariant name \ls`i` refined to be
associated with \ls`p`.
\begin{lstlisting}
let istore = list slprop $\qquad$ let istore_inv (i:istore) : slprop = List.fold_right ($\ast$) emp i $\qquad$ let inv_name = nat
let i_evolves is0 is1 = forall (i:inv_name). List.nth i is0 == None \/ List.nth i is0 == List.nth i is1
let inv_for_p (i:inv_name) (p:slprop) (m:mem) = Some p == List.nth i m.istore
let (~>) i p = witnessed (inv_for_p i p) $\qquad\qquad$ let ival (p:slprop) = i:inv_name{i~>p}

\end{lstlisting}

Now, to define the \ls`new_invariant p` action we
simply extend the \ls`istore`, witness that \ls`p` is now an
invariant, and return the address of the newly allocated invariant.
\begin{lstlisting}
let new_invariant (p:slprop) : action (ival p) p (fun _ -> emp) = fun () ->
  let m = get () in put ({m with istore=m.istore@[p]}); let i = List.length m.istore in witness (inv_for_p i p); i
\end{lstlisting}

With these definitions in place, we have all we need to instantiate
the \ls`state` interface of the semantics, using for each of its
fields (\ls`mem`, \ls`slprop`, \ls`evolves` etc.) the definitions
shown here.

\subsection{Using Invariants in Atomic Commands}
\label{sec:atomics}

We have seen how to allocate duplicable invariants, i.e., the analog
of the first two rules for manipulating invariants in Iris. What
remains is the third rule that allows invariants to be used in atomic
commands.

For starters, this requires carving out a subset of computations that
are deemed to be atomic, i.e., we need a way to express something like
the premise $\mbox{atomic(}e\mbox{)}$ from the Iris rule. However,
observe that our semantics from \S\ref{sec:semantics} already
provides a notion of atomicity: individual actions in \ls`Act` nodes
are run to completion without any interference from other threads.
\ch{My reading of the above is that actions {\bf must} be atomic.
  While reading below it seems that they {\bf may} be atomic. So quite confusing.
  Also is it really the memory model that defines actions?
  Yes, but I didn't get that at first.}%
Specific actions in our memory model can be marked as atomic,
depending on the particular architecture being modeled. For example,
one might include a primitive, atomic compare-and-set action, while
other primitive actions like reading, writing or allocating references
may or may not be atomic, depending on the architecture
being modeled. Further, some actions can be marked as \emph{ghost} and
sequences of such commands may also be considered atomic, since they
are never actually executed concretely.

Next, we need a way to determine which invariants are currently
``opened'' by an atomic command. Recursively opening the same
invariant \ls`ival p` is clearly unsound since, although \ls`ival p` is
duplicable, \ls`p` itself need not be.

Finally, to explain\ch{recover?}
Iris's atomic actions rule in full, we also need to model the \emph{later} modality \IrisNext. As we will
see, the \ls`witnessed` modality provided by the monotonic state effect serves
that purpose well.

\paragraph{The type of atomic actions} The type `\ls`atomic a uses is_ghost p q`'
below is a refinement\ch{not meant literally, but why not?}  of the
type of actions, \ls`action a p q` presented
in \S\ref{sec:sem-action}. The first additional index, \ls`uses`,
indicates the set of opened invariants---in particular, an atomic
action can only assume and preserve the invariants not included
in \ls`uses`, as shown in the definition of \ls`istore_inv'`. The
second index, \ls`is_ghost`, is a tag that indicates whether or not
this command is a ghost action. The \ls`atomic` type represents
an effectful operation a total sub-effect \ls`NMSTTot` of the effects
of nondeterminism and monotonic state---by choosing a total
sub-effect, we avoid pitfalls of infinitely opening invariants or
introducing divergence in ghost computations. As such, due to the
restriction to total computations, the type \ls`atomic a {} b p q` is
a subtype of the \ls`action a p q` type defined in the semantics.

\ch{Confused by the use of \ls`_` here.
  What exactly does it mean for something to be equivalent with \ls`_`?
  Is it equivalent to \ls`true` or to \ls`false`?}

\begin{lstlisting}
let istore_inv' uses ps = List.fold_right_i (fun p i q -> if i $\in$ uses then q else p `star` q) ps emp
let inv' uses m = $\ldots$ m.ctr $\ldots$ /\ istore_inv' uses m.istore
let atomic (a:Type) (uses:set inv_name) (is_ghost:bool) (p:slprop) (q: a -> slprop) =
    unit -> NMSTTot a mem mem_evolves
            (requires fun m -> interp (inv' uses m `star` p) m)
            (ensures fun m0 x m1 -> interp (inv' uses m1 `star` q x) m1 /\ preserves_frame p (q x) m0 m1)
\end{lstlisting}

We treat the \ls`atomic` type as a user-defined abstract effect in \fstar\ch{
  Where is that shown in use? A fwd reference might help. And what's the connection
  to the Steel effect in Section 5? Starting to get a bit lost.}
and insist on at most one non-ghost action in a sequential composition, as shown
by the signature of \ls`bind_atomic` below.

\begin{lstlisting}
val bind_atomic #a #b #u #p #q #r #g1 (#g2:bool{g1 || g2})
                (e1:atomic a u g1 p q) (e2: (x:a -> atomic b u g2 (q x) r)) : atomic b uses (g1 && g2) p r
\end{lstlisting}

\ch{The above is cool, but to what extent is it verified and to what extent just
  assumed/defined this way? Would \fstar{} complain if you would have gotten
  things wrong in bind\_atomic (say with respect to the gs) for instance?
  Where exactly? Is there a soundness proof of this in the style of Section 3.3?}

\paragraph{Opening and closing an invariant} The final piece of the puzzle
is the \ls`with_invariant` construct, whose signature is shown
below. Given an atomic command \ls`e` that uses the
invariant \ls`i:ival p` to gain and restore \ls`p`, it can be turned
into an atomic command that no longer uses \ls`i`, and whose use
of \ls`p` is no longer revealed in its specification.

\begin{lstlisting}
val with_invariant #a #p #q #r #u #g (i:ival p) (e:atomic a (i $\uplus$ u) g (p `star` q) (fun x -> p `star` r x)) : atomic a u g q r
\end{lstlisting}

Finally, given a value of type \ls`e:atomic a {} _ p q` we can promote
it to an \ls`e:action a p q` (since the types are equivalent)\ch{still confused
  by the \ls`_`; how can \ls`e:action a p q` be simultaneously equivalent to both
  \ls`e:atomic a {} true p q` and \ls`e:atomic a {} false p q`?}
and then turn it into a computation \ls`Act e : ctree a p q`.

\paragraph{See ya, later} The \ls`with_invariant` rule presented above
does not have Iris' \emph{later} modality, yet the later modality is
essential for soundness in Iris and in other
logics~\citep{dodds16sync} that support stored
propositions. Paraphrasing \citet{JungKJBBD18}, a logic that supports
allocating persistent propositions, together with a deduction rule for
the injectivity of stored propositions of the form
\ls@i ~> p `star` i ~> q $\vdash$ (p <==> q)@ is inconsistent---the
conclusion of the rule must be guarded under a later, i.e., it should
be \ls`$\IrisNext$(p<==>q)`. Although it may not be immediately
evident,~\citepos{preorders} model of monotonic state also has a
``later'' modality in disguise. In their model,
\ls`witnessed False $~\not\vdash~$ False`: instead, an explicit step
of computation via the \ls`recall` action is necessary to extract a
contradiction from \ls`witnessed False`. As such,
\ls@i ~> p `star` i ~> q $\vdash$ (p <==> q)@ is \emph{not} derivable
in \steelcore, although with a step of computation, the Hoare
triple
\ls@{i ~> p `star` i ~> q} recall i { p <==> q }@ is derivable.
In summary, the effect of monotonic state provides a way to account
for the necessary step indexing without making it explicit in the
logic.

\paragraph{The update modality and ghost actions}
As a final remark, allocating an invariant in Iris is done using
its \emph{update} modality, $\IrisUpdate$. Besides allocating
invariants, updates in Iris are also used to transform ghost
state. In \steelcore, rather than including such a modality within the
logic, we rely on \fstar's existing support for erased types to
model ghost state and updates within Hoare triples, rather than within
the logic itself.\footnote{Iris also internalizes Hoare triples, but
in \steelcore, we rely on the computation types of the host language
to express Hoare triples outside the logic.} For instance, the
following action represents a ghost read: it
dereferences \ls`x`, returning its contents only as an \ls`erased a`.

\begin{lstlisting}
val ghost_read (x:ref a p) : atomic (erased a) u true (exists v. pts_to r v) (fun v -> pts_to r v)
\end{lstlisting}

\subsection{Fractional Permissions and Monotonic References}
\label{sec:monotonic-refs}

Several prior works have provided PCM-based constructions both to
capture various sharing idioms as well as to define state machines
that constrain how the state is permitted to evolve. In this section,
we show how to use PCMs to encode~\citepos{preorders} preorder-indexed
monotonic references. We start, however, with a simpler construction
of references with fractional permissions, a construction we reuse for
monotonic references.

\paragraph{References with fractional permissions} To model references to
\ls`t`-typed values with fractional
permissions we store at each cell a value of type
\ls`frac t = option (t & r:real{0.0 < r})` with \ls`pcm_frac` as shown
below---the \ls`composable` predicate allows us to use undecidable
relations like propositional equality in our notion of partiality.

\begin{lstlisting}
let pcm_frac : pcm (frac t) = { one = None;
  composable = (fun f0 f1 -> match f0, f1 with | Some (v0, r0), Some(v1, r1) -> v0==v1 /\ r0+r1 <= 1.0 | _ -> True);
  op = (fun f0 f1 -> match f0, f1 with | None, f | f, None -> f | Some (v, r0), Some(_, r1) -> Some(v, r0 + r1)) }
\end{lstlisting}

Specializing the type of references and the points-to assertion for
use with \ls`pcm_frac`, we recover the traditional injective points-to
assertion on references and a lemma that relates the separating
conjunction in \ls`slprop` to composition in \ls`pcm_frac`.

\begin{lstlisting}
let ref t = ref (frac t) pcm_frac
let $(\mapsto_f)$ r v = pts_to r (Some (v, f)) `star` pure (f <= 1.0)
val share_gather (r:ref t) (f g:real) (u v:t) : Lemma (r $\mapsto_{f}$ u `star` r $\mapsto_{g}$ v) $\sim$ r $\mapsto_{f+g}$ u `star` pure (u==v))
\end{lstlisting}

\paragraph{Monotonic references} Whereas we have used preorders
and monotonic state within our memory model to support the dynamic
allocation of invariants, here we aim to expose preorders to describe
state transitions on individual references, in the style of Ahman et
al.'s monotonic references. Pleasantly, we find that our PCM-based
memory model layered above the monotonic state effect can precisely
capture Ahman et al.'s construction in a generic manner.

Our goal is to provide the following interface on an abstract
type \ls`mref a p` of references indexed by a preorder.  The main
point of interest is the signature of \ls`write`, which requires
proving that the new value \ls`v` is related to the old value by the
preorder \ls`p`.

\newcommand\mrefptsto{\ensuremath{\longmapsto}}

\begin{lstlisting}
val mref (a:Type) (p:preorder a) : Type
val $(\mrefptsto_{f})$ (x:mref a p) (v:a) : slprop
val read (r:mref a p) (v0:erased a) : action a (r $\mrefptsto_{f}$ v0) (fun v -> r $\mrefptsto_{f}$ v)
val write (r:mref a p) (v0:erased a) (v:a{p v0 v}) : action unit (r $\mrefptsto_{1.0}$ v0) (fun _ -> r $\mrefptsto_{1.0}$ v)
val observed (r:mref a p) (q:a -> prop) : prop
val witness_mref (r:mref a p) (q:stable_prop p) (v:erased a{q v})
    : action unit (pts_to r f v) (fun _ -> pts_to r f v `star` pure (observed r q))
val recall_mref (r:mref a p) (q:stable_prop a p) (v:erased a)
    : action unit (pts_to r f v `star` pure (observed r q) (fun _ -> pts_to r f v `star` pure (q v))
\end{lstlisting}

In return for respecting the preorder at each update, we provide two
new operations to witness and recall properties that are invariant
under the preorder. The operation \ls`witness` returns a pure,
abstract predicate \ls`observed r q` when the current value of \ls`r`
satisfies a stable property \ls`q`; and \ls`recall`
eliminates \ls`observed r q` into \ls`q v`, for \ls`v` the current
value of \ls`r`.
These operations are the analogue of the \ls`MST`
actions \ls`witness` and \ls`recall` exposed to \steelcore programs
at the granularity of a single reference, rather than the entire
state.
For instance, one could define a monotonically
increasing counter as \ls`r:mref int (<=)`, and having observed
that \ls`r` contains the value $17$ one can recall later that \ls`r`'s
value is at least $17$.

\paragraph{From PCMs to preorders} We observe that every
PCM induces a preorder and, dually, every preorder can be encoded as a
PCM. To interpret a PCM as a preorder, we take the infinite
conjunction of all preorders refined by the \ls`frame_preserving`
relation: in other words, since all updates must be frame-preserving,
we take the preorder of a PCM to be the strongest preorder entailed by
frame-preservation.
\begin{lstlisting}
let induces (p:pcm a) (q:preorder a) = forall (x y:a). frame_preserving p x y ==> (forall (z:a). p.compatible x z ==> q z y)
let preorder_of_pcm (#a: Type u#a) (p:pcm a) : preorder a = fun x y -> forall q. p `induces` q ==> q x y
\end{lstlisting}

With this notion in hand, we can finally define the heap evolution
relation (part of the global memory preorder shown
in \S\ref{sec:memory}) stating that (1) unused heap cells can change
arbitrarily; (2) used heap cells remain used; and, most importantly,
(3), the type and PCM associated with a ref cell does not change and
its value evolves according to the preorder of the PCM. In other
words, heaps evolve by the pointwise conjunction of the PCMs at each
ref cell.

\begin{lstlisting}
let h_evolves h0 h1 = forall (a:addr). match h0 a, h1 a with | None, _ -> True | Some _, None -> False
            | Some (Ref a0 p0 v0), Some (Ref a1 p1 v1) -> a0 == a1 /\ p0 == p1 /\ preorder_of_pcm p0 v0 v1
\end{lstlisting}

\paragraph{From preorders to PCMs}
Conversely, to interpret a preorder \ls`q:preorder a` as a PCM, we
define a PCM over \ls`hist q`, the type of histories over \ls`a`,
sequences of \ls`a`-values where adjacent values are related by \ls`q`,
with composability demanding one history to be an extension of the
other; composition being history extension; and the unit being the
empty history. The full construction is available in our online code
repository---we show its main signature below, including a round-trip
property, showing that the PCM built by the construction induces the
preorder corresponding to extension of \ls`q`-respecting histories.

\begin{lstlisting}
val pcm_of_preorder (q:preorder a) : p:pcm (hist q) {p `induces` history_extension}
\end{lstlisting}

This construction enables constructing a PCM \ls`frac_hist q` to
support the type \ls`mref a q`, combining fractional permissions with
the \ls`hist q` PCM, with the property that for any property
\ls`f:a -> prop` stable with respect to \ls`q`, its lifting
\ls`lift f : hist q -> prop` (that applies to the most recent
value in a history) is stable with respect to
\ls`preorder_of_pcm (frac_hist q)`. As such, the
underlying \ls$witness$ and \ls$recall$ operations of the
monotonic-state effect suffice to provide a model
for \ls`witness_mref`
and \ls`recall_mref`. In \S\ref{sec:examples:channel-types} we
use \ls`mref`s to encode a trace of messages exchanged on a channel,
proving that those traces respect a preorder induced by a
user-provided state machine.

\section{SteelCore at work: Locks, Fork/Join, Channels, Traces}
\label{sec:examples}

In this section, we make use of the \steelcore program logic to build
a few libraries of verified synchronization primitives. We start with
a spin lock, built using invariants accessed by an atomic CAS
instruction. Using a spin lock, we build a library for fork/join
concurrency on top of structural parallelism (\ls`par`)\ch{So far I've only
  seen \ls`Par`, and \ls`par` is only explained in 5.2, so wait?} and general
recursion. Finally, we present a library of synchronous, simplex
channels whose use is controlled by a specification-level state
machine. All of our examples are programmed directly within \fstar,
making use of all its abstraction and specification features,
including mixing dependently typed specifications and effects with
\steelcore's CSL. That said, while our proofs already rely on
both SMT solving and dependent typechecking, they are still
quite manual. As such, the code in this section is simplified to make it easier to explain: the full implementation together with the proofs are much more verbose
and are in the supplement. Better proof automation is left as future work.

\subsection{Example: Spin Locks}

We illustrate how to use invariants with atomic commands to build a
spin lock, building on a primitive compare-and-set atomic action with
the signature shown below. It states that given a (fractional
permission) reference \ls`r` to a word-sized integer for which we have
full permission, and \ls`old` and \ls`new` values, \ls`cas` updates
the reference to \ls`new` if its current value is \ls`old` and
otherwise leaves \ls`r` unchanged. Importantly, \ls`cas` is parametric
in the set of opened invariants \ls`u`. Note, \ls`cas` takes an
additional ghost parameter, \ls`v:erased uint32`, which represents the
value stored in the reference in the initial state.

\begin{lstlisting}
val cas (#u:set inv_name) (r:ref uint32) (old new:uint32) (v:erased uint32) :
    atomic (b:bool{b=(v=old)}) u false (r $\mapsto_{1.0}$ v) (fun b -> r $\mapsto_{1.0}$ (if b then new else v))
\end{lstlisting}

A lock is represented as a pair of a reference and an invariant
stating that the reference is in one of two states: either it holds
the value \ls`available` and the lock invariant \ls`p`, an
\ls`slprop`, is true separately; or it holds the
value \ls`locked`.

\begin{lstlisting}
let available, locked = false, true
let lockinv (r:ref bool) (p:slprop) = (pts_to r 1.0 available * p) \/ (pts_to r 1.0 locked)
let lock_t = ref bool & inv_name $\qquad$ let protects (l:lock_t) (p:slprop) : prop = snd l ~> lockinv (fst l) p
let lock p = l:lock_t { l `protects` p  }
\end{lstlisting}

For convenience, similarly to the NST effect in
\S\ref{sec:basic}, we package the \ls`ctree` trees into
a \ls`Steel` effect having the form \ls`Steel a fp0 fp1`.\ch{This is not
  true in the implementation though, which builds on top of the
  type of actions in a very shady way (par0 definition doesn't allow
  the 2 computations to be interleaved):
  \url{https://github.com/FStarLang/FStar/blob/steel/ulib/experimental/Steel.Effect.fst#L259}.}
Using this \ls`Steel` effect, allocating a lock is straightforward:

\begin{lstlisting}
let new_lock p : Steel (lock p) p (fun _ -> emp) =
    let r = alloc available in let i = new_invariant (lockinv r p) in (| r, i |)
\end{lstlisting}

Releasing a lock requires opening the invariant to gain permission to
the reference---we add comments to the code to show the relevant Hoare triples in the term using the
notation \ls`{p} e {q}`. Within the invariant, we
use a ghost read to fetch the current value of the reference, then do
a \ls`cas` and can prove that it sets the reference
to \ls`available`. In the case where the reference was already set to \ls`available`,
we use the affinity of our separation logic to forget the assertion
\ls`(b=false wand p)` before closing the invariant.
We could also return the resulting boolean to avoid losing information.

\begin{lstlisting}
let release ((| r, i |):lock p) : Steel unit p (fun _ -> emp) =
    let _ = with_invariant i
(*     {lockinv r p * p} *)
$\qquad$(*       {((pts_to r 1.0 available * p) \/ pts_to r 1.0 locked) * p} *)
         (let v = ghost_read r in cas r locked available v)
$\qquad$(*       {fun b -> pts_to r 1.0 available * (b=false wand p) * p} *)
(*    {lockinv r p * emp} *) in ()
\end{lstlisting}

Acquiring a lock is similar to releasing it: We try to set the
lock reference to \ls`locked` within the invariant using an atomic \ls`cas`.
If \ls`cas` fails, we "spin" by repeatedly calling \ls`acquire` until the lock becomes
available. The function terminates once the reference has been set to \ls`locked`
and we successfully acquired the corresponding \ls`slprop`.\ch{... or it loops
  forever if the lock is never released. There were various places in the paper
  in which you claimed that what you do wouldn't be too hard to do in other
  proof assistants, but this is not one of them. The fact that you could mark
  \ls`Steel` as a non-total effect seems to be crucial in this whole section;
  yet it's not even explicitly mentioned.}

\begin{lstlisting}
let rec acquire ((| r, i |):lock p) : Steel unit emp (fun _ -> p) =
  let b = with_invariant i
    (* {lockinv r p} *) (let v = ghost_read r in cas r available locked v) (* {fun b -> lockinv r p * (b=true wand p)} *)
  in if b then () else acquire (| r, i |)
\end{lstlisting}

Although \steelcore's logic is predicative, since the host
language supports abstraction in arbitrary universes, we can build
libraries whose specifications are generic in separation logic
assertions---this allows us to use our spin lock library to protect
any \ls`slprop`.

\subsection{Fork/Join}
\label{sec:forkjoin}

\steelcore's only concurrency primitive is the \ls`Par`
combinator for structured parallelism shown in
Figure~\ref{fig:m}, that we expose as a stateful \ls`par` in the \ls`Steel` effect.
However, having just built a library for locks, we
can code up a library for fork/join concurrency without too much
trouble. As with locks, since the host language is higher order, we
can easily abstract over computations and their specifications,
although Hoare triples are not part of \steelcore's logic itself.

The interface we provide for forking and joining threads is shown
below.
The type \ls`thread p` represents a handle to a thread which
guarantees \ls`p` upon termination.
The combinator \ls`fork f g` runs the thread \ls`f` and continues
with \ls`g` in parallel, passing to \ls`g` a handle to the thread
running \ls`f`.
The \ls`join t` combinator waits until the thread \ls`t` completes and
guarantees its postcondition.
\begin{lstlisting}
val thread (p:slprop) : Type
val fork #p #q #r #s (f: (unit -> Steel unit p (fun _ -> q))) (g: (thread q -> Steel unit r (fun _ -> s)))
  : Steel unit (p `star` r) (fun _ -> s)
val join #p (t:thread p) : Steel unit emp (fun _ -> p)
\end{lstlisting}

To implement this interface, we represent a thread handle as a boolean
reference protected by a lock that guarantees the thread's
postcondition \ls`p` when the reference is set. Allocating a thread
handle is easy, since the reference can initially be set to false.

\begin{lstlisting}
let thread p = { r:ref bool; l:lock (exists b. pts_to r 1.0 b `star` (if b then p else emp))}
val new_thread (p:slprop) : Steel (thread p) emp (fun _ -> emp)
\end{lstlisting}

To fork a thread, we create a new thread handle \ls`t`, then in
parallel, run \ls`g t` and in the thread for \ls`f`, we acquire the
lock, run \ls`f()`; then set the reference and release the lock.
\begin{lstlisting}
let fork #p #q #r #s f g =
  let t = new_thread q in let _ = par (fun _ -> acquire t.l; f(); t.r := true; release t.l) (fun _ -> g t) in ()
\end{lstlisting}

Finally, to \ls`join` we repeatedly acquire the lock, and if the
reference is set, we can free the reference and return the
postcondition \ls`p`; otherwise we release the lock and loop---\fstar's
existing support for general recursion makes it relatively easy.\ch{bingo,
  and that you're in a non-total effect}
\begin{lstlisting}
let rec join #p (t:thread p) = acquire t.l; let b = !t.r in if b then free t.r else (release t.l; join t)
\end{lstlisting}

Note, to provide a C-style fork/join on top of our API requires a
CPS-like transform, since \ls`fork` expects separate continuations for
the parent and child threads. We hope to address the usability of fork
in the future, perhaps layering another effect for continuations above
the Steel effect to support fork/join in direct style.

\subsection{Local State and Lock-Coupling Lists: Higher Order Assertions and Invariants}
\label{sec:higher-order}

Being embedded in a dependent type theory allows Steel programs to
enjoy all the abstraction facilities of the host language. In this
section, we provide two classic examples further illustrating the
abstraction facilities available, while pointing out some limitations.

\paragraph{Counters with local state} The type \ls`ctr_t` below represents
a counter as a closure paired with an abstract invariant \ls`p` over
its local state. The invariant is indexed by the current value of the
counter, and each application of the closure expects the invariant,
returns a value one greater than its previous index, and restores the
invariant at the returned value. To allocate a new
counter, \ls`new_ctr` returns a \ls`ctr_t` and provides the initial
invariant at the index \ls`0`. As mentioned in \S\ref{sec:slprop},
since \ls`slprops` cannot be stored in the heap, a \ls`ctr_t` cannot
be stored in a ref cell.

\begin{lstlisting}
let ctr_t = (p:(int -> slprop) & (x:erased int -> Steel (y:int{y==x+1}) (p x) p))
val new_ctr (_:unit) : Steel ctr_t emp (fun (| p, _|) -> p 0)
\end{lstlisting}

\paragraph{Lock-coupling lists} Spatial assertions and invariants can be
defined by recursion too, e.g., to describe the representation
invariant of a linked list, each of whose nodes is protected by its
own lock, a so-called lock-coupling list. A challenge here is to
support locks that can be dynamically allocated and stored in the heap
in alongside each node that it protects, but this is easily expressed
in our system, since invariants and locks (on which they are based),
are dynamically allocated and, unlike \ls`slprop`, are
storable. Proceeding along the lines of~\citet{gotsman07storable}, we
show the representation invariant for a lock-coupling list below.  The
predicate \ls`llist_inv repr n` grants ownership to the head of the
list (if any); whose value \ls`v` validates \ls`p`; and states that
the lock stored at the head recursively grants the representation
predicate for the tail of the list.

\begin{lstlisting}
type llist (a:Type0) : Type0 = {  v : a;  next : ref (llist a);  lock : lock_t }
let rec llist_inv (repr:list (a -> slprop)) (n:ref (llist a)) =
  match repr with | [] -> emp | p::tl -> exists c.  p c `star` n $\mapsto_{1.0}$ c `star` pure (c.lock `protects` llist_inv tl c.next)
\end{lstlisting}

\subsection{Channel Types: From Indexed Action Trees to Action Tree Indexes}
\label{sec:examples:channel-types}

As a final example, we present a library for synchronous communication
among threads. We draw on inspiration from the long line of work on
session types~\citep{honda98sessions} to enforce a typing discipline
that associates with each channel a state machine that describes the
sequence of permissible operations on that channel.
For this paper, we focus only on the simplest scenario of synchronous
simplex channels (channels with unidirectional
communication)---towards the end, we remark on how to generalize this
construction to 2-party session types and duplex channels. As such,
our work shows how to use \steelcore as a platform on which to model
higher-level constructs for reasoning about concurrent and distributed
programs, while mixing various concurrency idioms (e.g., channels,
locks and atomics).

Our model of channels proceeds in three steps. First, we define a
small language for describing protocols as action trees. Next, we
define a notion of partial traces of protocols, sequences of messages
that are accepted by the protocol state machine. And, finally, we
define the type of channels indexed by protocols with an interface
that supports sending and receiving messages on channels, together
with an internalized proof (done once and for all protocol-indexed
channels) that the sequence of messages received on a channel are a
partial trace of the protocol.

\subsubsection{Protocols as Action Trees}

The type \ls`protocol` below expresses a small embedded language to
express the sequence of messages that can be sent on a
channel---the \ls`erasable` annotation causes \fstar to check that the
type is never used in a computationally relevant context and
all \ls`protocol` values are erased to \ls`()` during extraction.
\begin{lstlisting}
[@erasable] type protocol : Type -> Type =
| Ret     : #a:Type -> v:a -> protocol a
| Msg     : a:Type -> #b:Type -> k:(a -> protocol b) -> protocol b
| DoWhile : p:protocol bool{p<>Ret _} -> #a:Type -> k:protocol a -> protocol a
\end{lstlisting}
The type has the classic structure of an infinitely branching tree of
actions similar to the one described in \S\ref{sec:background}.
For example, the term below describes a small two message protocol,
where the first message is an integer \ls`x` and the second
message \ls`y` is an integer one greater than the first message, i.e.,
the continuations of a message depend on the values exchanged in the
history of the protocol.
\begin{lstlisting}
let xy = Msg int (fun x -> Msg (y:int{y = x + 1}) (fun y -> Ret ()))
\end{lstlisting}
It should be straightforward to see that \ls`protocol` is a monad,
with an easily definable \ls`bind`. The \ls`DoWhile` construct allows
specifying infinite protocols. For example,
\ls$DoWhile (xy `bind` (fun _ -> Ret true)) (Ret ())$ is a channel
on which one can repeatedly send related pairs of successive
integers.

\subsubsection{Traces of a Protocol}
One may wonder how we give semantics to infinite reductions
with \ls`DoWhile`: as it turns out, we only consider finite partial
traces defined as the reflexive, transitive closure of a single step
relation, as we show next.

The function \ls`hnf` below puts a protocol into a form that begins
with either a \ls`Ret` (in case the protocol has finished) or
a \ls`Msg` node, indicating the next action to be performed.
\begin{lstlisting}
let rec hnf (p:protocol a): (q:protocol a{(Ret? q \/ Msg? q) /\ (~(DoWhile? p) ==> (p == q))})
  = match p with | DoWhile p k -> bind (hnf p) (fun b -> if b then DoWhile p k else k) | _ -> p
\end{lstlisting}

Using it, we can define a notion of a single step of reduction of a
protocol: if a protocol \ls`p` has more actions, then given a
message \ls`x` whose type matches the type of the next action, \ls`p`
steps according to its continuation.
\begin{lstlisting}
let more (p:protocol a) : bool = Msg? (hnf p)
let msg_t (p:protocol a) : Type = match hnf p with | Msg a _ -> a | Ret #a _ -> a
let step (p:protocol a{more p}) (x:msg_t p) : protocol a = Msg?.k (hnf p) x
\end{lstlisting}

The type of traces below, \ls`trace from to`, represents a sequence of
messages that are related by stepping the protocol \ls`from` until the
protocol \ls`to`, and \ls`trace_of p` is the type of traces accepted
by zero or more steps of \ls`p`.
\begin{lstlisting}
let prot = protocol unit
[@erasable] type trace : prot -> prot -> Type = | Waiting  : p:prot -> trace p p
  | Message  : from:prot{more from} -> x:msg_t from -> to:prot -> trace (step from x) to -> trace from to
type trace_of p = { until:prot;  tr:trace p until }
\end{lstlisting}

It's easy to extend a trace by one message; then to define a relation
\ls`next` on t; and finally to define trace extension as the closure
of \ls`next`.
\begin{lstlisting}
val extend_1 (#from #to:protocol unit) (t:trace from to{more to}) (m:msg_t to) : trace from (step to m)
let next p t0 t1 = more t0.to /\ (exists msg. t1.to == step t0.to msg /\ t1.tr == extend_1 t0.tr msg)
let ($\hookrightarrow$) (#p:protocol unit) : preorder (trace_of p) = ReflexiveTransitiveClosure.closure (next p)
\end{lstlisting}

We will use the preorder $\hookrightarrow$ to maintain a ghost monotonic
reference storing a log of messages associated with a channel and to
prove that the trace of messages on a channel are always accepted by
that channel's protocol.

\subsubsection{Channel Types}
\label{sec:channels}

We aim to provide the following (hopefully idiomatic) interface to
work with channels. The abstract type \ls`chan i` is the
type of channels created for use with the \emph{initial
protocol} \ls`i`. We have two abstract predicates, \ls`sender`
and \ls`receiver`, both indexed by a protocol that describes
the \emph{current state} of the channel from the sender's and
receiver's perspective, respectively

\begin{lstlisting}
val chan (i:prot)  : Type
val sender #i (c:chan i) (cur:prot) : slprop
val receiver #i (c:chan i) (cur:prot) : slprop
\end{lstlisting}

To create a channel, we use \ls`new_chan i`, we return a new \ls`chan`
and the sender's and receiver's state separately initialized to the
given initial protocol \ls`i`.

\begin{lstlisting}
val new_chan (i:prot) : Steel (chan i) emp (fun c -> sender c i * receiver c i)
\end{lstlisting}

To \ls`send` a message, one presents a channel \ls`c` in a state
\ls`cur` where more messages are expected; a value \ls`x`, whose type is
the type of the next message in the protocol: as a result, the
sender's state transitions by a single step, which depends on the
value \ls`x` provided. The \ls`recv` is dual to the send.

\begin{lstlisting}
val send #i (#cur:prot{more cur}) (c:chan i) (x:msg_t cur) : Steel unit (sender c cur) (fun _ -> sender c (step cur x))
val recv #i (#cur:prot{more cur}) (c:chan i) : Steel (msg_t cur) (receiver c cur) (fun x -> receiver c (step cur x))
\end{lstlisting}

In addition, we provide further operations that internalize the
guarantee that the trace of messages on a channel respects its
protocol.  The duplicable abstract predicate \ls`history c t` states
that \ls`t` is a partial trace of messages received on \ls`c`. The
operation \ls`trace` allows a client to extract the current
trace. Most importantly, \ls`extend_trace` ensures that traces are
monotonic: if \ls`history c p` witnesses that \ls`p` was a trace
of \ls`c`, then if the receiver's current state is \ls`cur`, one can
prove that the current trace \ls`t` of the protocol is an extension
of \ls`p` until \ls`cur`. That is, all well-typed channel programs
respect the channel's protocol.
\begin{lstlisting}
val history #i (c:chan i) (t:trace_of i) : slprop
val history_duplicable #i (c:chan i) (t:trace_of i) : Steel unit (history c t) (fun _ -> history c t `star` history c t)
val trace #i (c:chan i) : Steel (trace_of i) emp (fun tr -> history c tr)
val extend_trace #i (#cur:prot) (c:chan i) (p:trace_of i): Steel (t:trace_of i{p `extended_to` t})
            (receiver c cur `star` history c p) (fun t -> receiver c cur `star` history c t `star` until t == cur)
\end{lstlisting}

Implementing this interface will take a few steps, and will exercise
nearly all elements of SteelCore presented so far, including
fractional permissions, ghost state, locks, and monotonic references.
\ch{Missing dependency to monotonic references in Fig 1}

\paragraph{Representing channels}

We'll represent\ch{+synchronous} channels by a pair of concrete references, each
writable by only one side of the channel, though readable by both (via
a lock), and a ghost reference maintaining a trace of the protocol.
The concrete references contain a triple: an erased field \ls`prot`, which
we'll use to state our invariants; the last value sent or received on
the channel (respectively); and a counter \ls`nat` which counts the
number of messages sent or received so far, which we'll use to
determine if a message is available to be received or not.
\begin{lstlisting}
type chan_val = { prot : prot{more prot}; msg  : msg_t prot; ctr : nat}
type chan_t i = { send : ref chan_val; recv : ref chan_val; trace : mref (trace_of i) ($\hookrightarrow$) }
\end{lstlisting}

The main invariant \ls`chan_inv` retains half a permission on the
concrete references and full permission on the trace. It states that
the trace is a partial trace of \ls`i` until the state of the
protocol on the receiver's side. Finally, it states that either the
last sent message has already been received, in which case, the
contents of the two references agree. Or, the sender is exactly one
step of the protocol ahead of the receiving reference---its counter is
one greater, and its protocol is a single step ahead of the receiver.
\begin{lstlisting}
let chan_inv #i (c:chan_t i) = exists vs vr tr. pts_to c.send 0.5 vs `star` pts_to c.recv 0.5 vr `star` pts_to c.trace 1.0 tr `star`
  (tr.until == step vr.prot vr.msg) `star`
  (if vs.chan_ctr = vr.chan_ctr then vs==vr else vs.ctr==vr.ctr+1 /\ vs.prot == step vr.prot vr.msg )
\end{lstlisting}

Our channel type packages the two references with a lock that protects
the \ls`chan_inv` invariant. This makes channels fully first-class:
channels can be stored in the memory and channels can even be passed
on channels.
\begin{lstlisting}
let chan i = { chan : chan_t i; lock : lock (chan_inv chan) }
\end{lstlisting}

The \ls`sender c p` predicate is a permission to transition the
protocol by one step to state \ls`p`:
it retains half a permission to the
\ls`c.chan.send` reference, together with an assertion that \ls`p` is
exactly the successor state of the protocol stored in the reference.
The \ls`receiver` predicate is
similar. In both cases, by retaining half a permission to the
reference, acquiring the \ls`chan_inv` lock gives a full permission to
the reference in question allowing, but only allowing
the other reference to be read.
\begin{lstlisting}
let sender (c:chan) (p:prot) = exists vs. p == step vs.prot vs.msg /\ pts_to c.chan.send 0.5 vs
let receiver (c:chan) (p:prot) = exists vr. p == step vr.prot vr.msg /\ pts_to c.chan.recv 0.5 vr
\end{lstlisting}

\paragraph{Implementing channels} With these invariants in place,
the implementation is nearly determined. For space reasons, we only
sketch the implementations of \ls`recv` and \ls`extend_trace`.

Receiving a message involves acquiring the lock, reading both
references, and if a message is not available, releasing the lock and
looping; otherwise, we update the receiver's state, extending their
trace (ghostly), releasing the lock and returning the received
message.
\begin{lstlisting}
let rec recv #i #cur c = acquire c.lock; let vs, vr = !c.chan.send, !c.chan.recv in
    if vs.ctr=vr.ctr then release c.lock; recv c
    else c.chan.recv := vs; c.chan.trace := extend_1 c.chan.trace vs.msg; release c.lock; vs.msg
\end{lstlisting}

Witnessing the monotonicity of the trace makes use of the monotonic
references from \S\ref{sec:monotonic-refs}. We define the \ls`history`
predicate in terms of the observations on monotonic
references. Extending a trace then involves acquiring the lock,
reading the trace, recalling the prior observation to learn that the
current trace is an extension of the previous one, then making another
observation using \ls`witness_mref`; and finally releasing the lock and returning.

\begin{lstlisting}
let history c tr = observed c.chan.trace (tr $\hookrightarrow$ _)
let extend_trace #i #cur c prev = acquire c.lock; let tr = !c.chan.trace in recall_mref c (prev $\hookrightarrow$ _);
                              witness_mref c.chan.trace (tr $\hookrightarrow$ _) tr; release c.lock; tr
\end{lstlisting}

\paragraph{2-party sessions} Channel types are already
a useful abstraction, but they would be even more so when generalized
to support duplex, asynchronous channels. We do not foresee any major
difficulties in doing so. To support asynchrony, rather than holding
just a single message in the sender's reference, we can buffer
messages in the sender's state and the receiver can dequeue them. To
support duplex channels, we anticipate extending the language of
protocols to support directed message actions between principals and
to derive mutually dual protocols for each participant by inverting
the polarities of each messaging action. We leave both of these
extensions to future work.

\paragraph{Discussion}
In a sense, we've come full circle: we started
in \S\ref{sec:semantics} by representing infinite concurrent
computations as indexed effectful action trees. Now, we specify
concurrent programs using indexed types, where the indexes themselves
are action trees with infinite traces, with a proof within \steelcore
that the message traces are partial traces of the index state
machines. To prove safety properties of concurrent, channel-using
programs, one can reason by induction on the partial
traces. Alternatively, rather than reason directly on traces, one
might even replay the methodology of this paper ``one level up'' and
derive a program logic to reason about action tree indexes.
\ch{Any intuition why this is easier, and not just reproducing
  the same problems one level up?}

\section{Related work}
\label{sec:related}

Throughout the paper, we have discussed connections to many strands of
related work on CSL. In particular, we have
drawn inspiration from, and contrasted our work with,
Iris~\citep{JungKJBBD18}. The most significant point of contrast with
Iris, perhaps, is our differing goals. Iris is a powerful
impredicative logical framework into which other logics and
programming languages can be embedded and studied. This has allowed
researchers to use Iris as a foundation on which to investigate
languages and language features as different as unsafe blocks in
Rust~\citep{jung17rustbelt} and state-hiding via rank-2 polymorphism
in Haskell~\citep{timany18runst}. In contrast,
with \steelcore, we aim not to provide a general logical framework but
instead to extend a proof assistant's programming language with an
effect for concurrency and to reason about effectful, dependently
typed concurrent programs in a CSL. This
allows us to keep the embedded logic relatively simple: unlike Iris,
it does not internalize Hoare triples, does not support storing \ls`slprop`s, and does not
make use of step indexing (since our model internalizes monotonicity) or any of Iris' several modalities. However,
many lacking features in the logic are recovered using the facilities of the host
language, \fstar. E.g., we support rules for
manipulating dynamically created, named invariants \ls$i ~> p$ in a
style inspired by Iris, using the underlying
effect of monotonic state. Further, the logic supports quantification over \ls`slprop`s
while remaining in \fstar's predicative universe hierarchy, and also supports abstraction
at the level of \fstar's programming language (\S\ref{sec:examples}), e.g., we can
abstract over \ls`slprop`s, computations with Hoare triples, etc.

In developing a shallow embedding of CSL in a dependent type
theory, \steelcore is similar to
FCSL~\citep{nanevski14fcsl,sergey15fcsl,nanevski19fcsl}. FCSL is
shallowly embedded in Coq and relies on Coq's abstraction facilities
for some of its expressive power. Their logic (like ours and unlike
Iris') applies directly to Coq programs, rather than to embedded
programs. FCSL's semantic model is also similar to ours, in that they
also represent computations as action trees. However, rather than
using indexed action trees and directly interpreting the trees as the
proof rules of a Hoare logic, Nanevski et
al. (like~\citepos{brookes04csl} original proof of soundness of CSL)
instead go via the indirection of action traces. In principle, it
might be possible to define something like FCSL's logic using indexed
action trees and to interpret those trees directly into the Hoare Type
Theory underlying FCSL. However, modeling partiality in this way
within Coq may be difficult, if not impossible---the action traces
approach provides a way around this. In contrast, working in
\fstar's effectful type theory, we can directly model partiality
and avoid the indirection of traces. FCSL's action traces also
resemble the trace semantics we developed for our action trace indexes
for channel types (\S\ref{sec:channels}). Another notable point of
distinction with FCSL is \steelcore's treatment of invariants.
Since FCSL's model is predicative, it does not provide a way to dynamically
allocate an invariant,
making it impossible to model certain kinds of
synchronization primitives, e.g., our generic interface for locks does
not seem to be expressible in FCSL. On the other hand, FCSL provides
several constructs for reasoning about concurrent programs mixing
styles of reasoning from CSL with rely-guarantee reasoning, something
which we haven't explored much: our use of monotonic references may
play a role in this direction, particularly in connection with other
related work on rely-guarantee references~\citep{gordon13rgref}.

Many researchers have explored using action trees in modeling
effectful constructs, including building indexed representations and
interpreting them into effectful
computations~\citep{Mcbride11kleisliarrows,brady13effects}. Allowing
effects in constructors of inductive types has been separately studied
for simple, non-indexed types by \citet{filinski07effectfuldata}
and \citet{atkey15interleaving}. We are the first to consider indexed
effectful action trees that mix data and effectful computations, while
interpreting the trees into another indexed effect, allowing us to
layer effects---in our case, layering concurrency over divergence,
monotonic state, and nondeterminism---while also deriving a program
logic to reason about the new effect layer, based on the indexing
structure.  Our action trees from Section~\ref{sec:semantics} have one
parameter (the state type class) and four logical specification (including implicit dynamic frames)
indexes in addition to the result type.
This layering of effects also allows us to support infinite
computations, without needing coinduction. In
contrast,~\citet{xia20interaction-trees} build coinductive,
non-indexed action trees and give them an extrinsic, equational
semantics. It would be interesting to study whether our style of
intrinsically defined program logics can be developed using indexed
versions of Xia et al.'s coinductive action trees.

We prove the soundness of our semantics by building an intrinsically
typed definitional interpreter for our indexed effectful action
trees. Intrinsically typed definitional interpreters have been
investigated before by~\citet{poulsen17definitional}, who give several
instances for languages ranging from the simply typed lambda calculus
to middleweight Java embedded in Agda. The typing guarantees from
these definitional interpreters ensure syntactic type safety of
reduction with respect to the classic type systems for the languages
they study. More recently, ~\citet{rouvoet20linear} give an
intrinsically typed definitional interpreter for a linearly typed
language, which bears some resemblance (owing to its linearity) to the
structure of our interpreter for CSL. Our approach is similar to both
these works in spirit, with two notable differences. First, rather
than defining interpreters for deeply embedded languages, our
representation allows the interpretation of host-language terms in an extended
effectful semantics including concurrency (following the free monads
methodology of~\citet{swierstra2008alacarte} and others). Second,
rather than proving syntactic type safety for embedded programs, we
derive the soundness of a generic concurrent separation logic applied
to host programs.

Embedding session types in CSL has also been investigated before---the
Actris system embedded in Iris explores this in
depth~\cite{hinrichsen19actris}. Our channel types explore a similar
direction, though we only scratch the surface, in that\ch{phrasing} as a
proof concept of the expressiveness of the underlying logic, we only
give an encoding for synchronous simplex channels. One technical
difference is that we embrace the use of state transition systems
structured as action trees and prove trace inclusion within the system
once and for all. Actris instead provides impredicative dependent
separation protocols, which due to the use of higher-order ghost state
that can depend on the type of propositions, appear to be
strictly more expressive than our predicative state
machines. Nevertheless, both systems can describe data-dependent
protocols. Prior work on \fstar~\citep{SwamyCFSBY11}, which then
included support for affine types, also developed a model for
data-dependent affinely typed sessions  for sequential,
pure \fstar programs. Nearly a decade later, we show how to encode
channel types for concurrent, stateful \fstar programs with CSL.
\ch{Another full circle is with respect to monotonic state appearing in the form
  of a monotonic log (soup) of propositions in the process calculus world of
  RCF/F7~\cite{BengtsonBFGM11}, then being ported to the sequential world of
  \fstar{}~\cite{preorders}, and now being a key ingredient for supporting
  concurrency/distribution again.}

\section{Conclusion}
\label{sec:conclusions}

We have demonstrated how a full-fledged CSL can be embedded in an
effectful dependent type theory, relying on an underlying semantics of
monotonic state to model features that have otherwise required
impredicative logics.
In doing so, we have brought together two strands of work pioneered by
John Reynolds: definitional interpreters and separation logic---we
hope that he would at least have been intrigued by our work.
Going forward, we plan to make use of \steelcore as the foundation of
a higher level DSL embedded in \fstar, aiming to use a combination of
tactics for manipulating \ls`slprop`s and SMT solving for implicit
dynamic frames to help ease proofs of concurrent programs.

\ifanon
\else
\section*{Acknowledgments}
Aymeric Fromherz's and Denis Merigoux's work was supported in part by
internships at Microsoft Research. Aymeric Fromherz was also funded by
the Department of the Navy, Office of Naval Research under Grant
no. N00014-18-1-2892. Denis Merigoux was also funded by ERC
Consolidator Grant CIRCUS no. 683032. Danel Ahman's work was supported
in part by the Microsoft Research Visiting Researcher program. He has
also received funding from the European Union’s Horizon 2020 research
and innovation programme under the Marie Skłodowska-Curie grant
agreement no. 834146. We thank Robbert Krebbers, the shepherd of this
paper; the anonymous reviewers; Matt Parkinson, Derek Dreyer, Ralf
Jung, Aleks Nanevksi and all the members of Project Everest for their
feedback, answering questions and many useful discussions.
\fi

\bibliographystyle{abbrvnaturl}
\bibliography{fstar,steel}
\end{document}